\documentclass[aps,prb,reprint,twoside,showpacs,floatfix,superscriptaddress]{revtex4-1}
\usepackage{amssymb}
\usepackage{amsmath}
\usepackage{graphicx}
\usepackage{times}
\usepackage{xcolor}
\usepackage[colorlinks=true,linkcolor=blue,citecolor=blue,bookmarks=false]{hyperref}
\usepackage{CJK}

\input{tcilatex}
\begin{document}

\begin{CJK*}{Bg5}{bsmi}
\title{Spin and charge transport in U-shaped one-dimensional channels with spin-orbit couplings}
\author{Ming-Hao Liu (¼B©ú»¨)}
\email{Ming-Hao.Liu@physik.uni-regensburg.de}%
\affiliation{Institut f\"{u}r Theoretische Physik, Universit\"{a}t Regensburg, D-93040 Regensburg, Germany}%
\affiliation{Department of Physics, National Taiwan University, Taipei 10617, Taiwan}%
\author{Jhih-Sheng Wu (§d­P²±)}
\affiliation{Department of Physics, National Taiwan University, Taipei 10617, Taiwan}%
\author{Son-Hsien Chen (³¯ªQ½å)}
\affiliation{Department of Physics, National Taiwan University, Taipei 10617, Taiwan}%
\affiliation{Department of Physics, Georgetown University, 37th and O Sts.\ NW, Washington, D.C.\ 20057, USA}%
\author{Ching-Ray Chang (±i¼y·ç)}
\email{crchang@phys.ntu.edu.tw}
\affiliation{Department of Physics, National Taiwan University, Taipei 10617, Taiwan}%
\begin{abstract}%
A general form of the Hamiltonian for electrons confined to a curved
one-dimensional (1D) channel with spin-orbit coupling (SOC) linear
in momentum is rederived and is applied to a U-shaped channel.
Discretizing the derived continuous 1D Hamiltonian to a
tight-binding version, the Landauer--Keldysh formalism (LKF) for
nonequilibrium transport can be applied. Spin transport through the
U-channel based on the LKF is compared with previous quantum mechanical
approaches. The role of a curvature-induced geometric potential which
was previously neglected in the literature of the ring issue is also
revisited. Transport regimes between nonadiabatic, corresponding to
weak SOC or sharp turn, and adiabatic, corresponding to strong SOC
or smooth turn, is discussed. Based on the LKF, interesting charge and
spin transport properties are further revealed. For the charge
transport, the interplay between the Rashba and the linear
Dresselhaus (001) SOCs leads to an additional modulation to the
local charge density in the half-ring part of the U-channel, which
is shown to originate from the angle-dependent spin-orbit potential.
For the spin transport, theoretically predicted eigenstates of the
Rashba rings, Dresselhaus rings, and the persistent spin-helix
state are numerically tested by the present quantum transport
calculation.
\end{abstract}
\date{\today}
\maketitle
\end{CJK*}

\section{Introduction}

Recent progress in the experimental techniques fabricating semiconductor
nanostructures\cite{Ihn2010} has made low-dimensional electronic transport
one of the enduring focuses in condensed-matter physics. For one-dimensional
(1D) systems, quantum wires (QWs) can be realized by growing nanowires such
as semiconductor-based nanowhiskers or carbon nanotubes. In layered
semiconductors, formation of QWs by confining the electron gas to a quasi-1D
region is also possible in various ways, such as V-groove quantum wells,
cleaved-edge overgrowth, or atomic force microscopy (AFM) lithography.\cite%
{Fuhrer2002} The latter provides an even more flexible way of designing the
shape of the confinement, and a quantum ring (QR) is one of the important
examples. 
\begin{figure}[b]
\centering\includegraphics[width=\columnwidth]{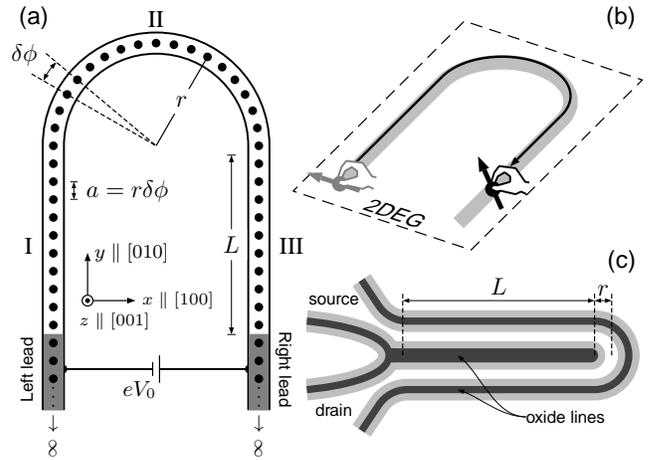}
\caption{(a) Schematic sketch of the tight-binding model for the U-channel.
Left and right arms are labeled as regions I and III, respectively, both
with $N_{w}$ sites. Region II is the half-ring part with $N_{r}$ sites. (b)
In quantum mechanical approaches, the electron spin propagates through the
U-channel via either the translation operator or the spin propagator, which
is virtually identical to dragging the spins by hand along a U-shaped path
in two dimensions. (c) Schematic sketch of the U-channel realized by AFM
lithography. }
\label{fig1}
\end{figure}

1D transport in QWs was previously focused on the charge properties.\cite%
{Ferry2009} A subsequent intensive investigation on spin-dependent transport
was triggered ever since the proposal of the Datta--Das transistor,\cite%
{Datta1990} whose underlying mechanism is based on the Rashba spin-orbit
coupling (SOC) due to structural inversion asymmetry.\cite{Bychkov1984} On
the other hand, QRs provide a natural platform to study the Aharonov--Bohm
effect\cite{Aharonov1959} in solids. The idea of a \textquotedblleft
textured\textquotedblright\ magnetic field applied on the QR\cite{Loss1990}
opened the study of the Berry phase\cite{Berry1984} in rings, in which the
adiabatic transport plays a key role. The Berry phase acquired by the
electron spin in rings was later discussed,\cite{Stern1992} and
investigation of the Rashba effect in QRs was subsequently initiated,\cite%
{Aronov1993} although the employed Hamiltonian at that time was
\textquotedblleft incorrect.\textquotedblright\ After the \textquotedblleft
correct\textquotedblright\ ring version of the Rashba Hamiltonian was
derived by Meijer \textit{et al.} almost a decade later,\cite{Meijer2002} a
series of theoretical discussion over the Rashba ring issue continued until
recently.\cite%
{Splettstoesser2003,Frustaglia2004,Molnar2004,Souma2004,Foldi2005,Wang2005,Foldi2006}

So far we have been reviewing planar 1D systems where the curvature either
vanishes (QWs) or globally exists (QRs), whereas a general 1D system may
include a position dependent curvature. Quantum mechanical particle motion
confined to a surface was first discussed by Jensen and Koppe\cite%
{Jensen1971} and da Costa,\cite{Costa1981} regardless of spin, and was later
generalized to include the SOC effect.\cite{Entin2001} When further
restricted to a curved planar 1D wire, da Costa proposed a linear potential
term due to curvature,\cite{Costa1981} which was later termed the \emph{%
geometric potential} and has recently been observed in photonic crystals.%
\cite{Szameit2010}

Spin transport in a curved 1D wire in the presence of SOC was recently
discussed.\cite{Trushin2006,Liu2006b,Zhang2007} In Ref.\ %
\onlinecite{Trushin2006}, however, only the Rashba SOC was considered; in
Ref.\ \onlinecite{Liu2006b}, both Rashba SOC and the Dresselhaus (001)
linear term (arising from the bulk inversion asymmetry of the underlying
crystal\cite{Dresselhaus1955}) were taken into account, but both Refs.\ %
\onlinecite{Trushin2006} and \onlinecite{Liu2006b} did not consider the
geometric potential. In Ref.\ \onlinecite{Zhang2007}, the geometric
potential was considered but the SOC included only the Rashba term. Hence a
more complete study of the spin transport in curved 1D wires, taking into
account both Rashba and Dresselhaus terms, as well as the curvature-induced
geometric potential, is essential.

Regardless of the geometric potential, spin precession due to SOCs along
arbitrary paths was previously studied quantum mechanically, either through
the conventional translation operator\cite{Liu2006b} or a spin propagator
obtained by a properly defined spin-orbit gauge.\cite{Chen2008a} Despite the
fact that the electron spin is indeed forced to evolve through a 1D path,
these spin precession studies\cite{Liu2006b,Chen2008a} are built on a
two-dimensional nature--there is no confinement. Thus how well this simple
quantum mechanical picture can survive when a more realistic situation is
considered, such as a lead-conductor system subject to electric bias, has
for years been a question we would like to answer.

In this paper, spin precession patterns along a curved 1D wire based on the
previous formalisms, namely, quantum mechanical space translation [and its
approximating result spin vector formula (SVF)]\cite{Liu2006b} and
spin-orbit gauge method\cite{Chen2008a} will be qualitatively and
quantitatively compared with those obtained by the more sophisticated
nonequilibrium Green's function formalism\cite{Datta1995} [or in ballistic
systems free of particle-particle interaction, the Landauer--Keldysh
formalism (LKF)\cite{Nikolic2005a,Nikolic2006}]. Meanwhile, we will
reinvestigate the influence of the geometric potential in curved 1D
transport. These are regarded as our first goal. Whereas the SOCs in general
depend on the momentum, electron spin traversing a curved 1D wire encounters
a varying effective magnetic field. This resembles the textured magnetic
field\cite{Loss1990} and is therefore closely related to the issue of
adiabatic transport, which is our second goal in the present paper.

For these purposes we consider a U-shaped 1D channel, composed of two
straight QWs and a half-QR in between, and theoretically inject electron
spin from the source end and analyze the spin orientation along the
U-channel down to the drain end. For computational concern, the U-channel is
descretized into a finite number of lattice grid points, as sketched in
Fig.\ \ref{fig1}(a). We label the left and right QWs of the U-channel as
regions I and III, respectively, each containing $N_{w}$ sites, and the
half-ring as region II, containing $N_{r}$ sites. In addition to the listed
two goals, further investigation of the charge and spin transport properties
based on the LKF will be the last goal. For the charge transport, the
interplay between the Rashba and the linear Dresselhaus (001) SOCs leads to
an additional modulation to the local charge density in the half-ring part
of the U-channel and will be shown to originate from the emergence of the
angle-dependent spin-orbit potential. For the spin transport, theoretically
predicted eigenstates of the Rashba rings,\cite%
{Splettstoesser2003,Molnar2004,Frustaglia2004,Foldi2005,Wang2005}
Dresselhaus rings,\cite{Wang2005} and the persistent spin-helix state\cite%
{Bernevig2006,Liu2006c,Koralek2009} are numerically tested by the present
quantum transport calculation.

This paper is organized as follows. In Sec.\ \ref{sec theory} we introduce
the Hamiltonians and briefly the formalisms to be used in the transport
calculations, which are reported next in Sec.\ \ref{sec transport analysis}.
Numerical results carrying out the above-listed three goals are reported,
respectively, in Secs.\ \ref{sec QMvsQT}, \ref{sec spin flip}, and \ref{sec
charge and spin transport}. Experimental aspects regarding the fabrication
of the U-channel are given in Sec.\ \ref{sec exp}. We conclude in Sec.\ \ref%
{sec conclusion}.

\section{Theory\label{sec theory}}

In this section we will introduce the Hamiltonians to be used in the LKF,
and review the different theoretical approaches for the spin transport
calculation.

\subsection{Hamiltonians}

In the following we first review and rederive the general form of the
Hamiltonian for a continuous curved 1D system, and then apply it to the 1D
ring case, which is the nontrivial part of our U-channel. We then write its
corresponding tight-binding version of the Hamiltonian, to be used in the
LKF calculation. Throughout we will not explicitly discuss the Hamiltonian
for the straight parts of the U-channel since they are relatively trivial
and well known.

\subsubsection{Continuous curved 1D systems: General form}

Consider the motion of electrons confined in a 1D planar curvilinear wire.
Electrons originally in a two-dimensional plane are confined to a quasi-1D
channel. We will derive the 1D effective Hamiltonian in the presence of
SOCs. Under the effective-mass approximation in solids, the Hamiltonian for
an electron in our model is 
\begin{equation}
\mathcal{H}=\frac{\mathbf{p}^{2}}{2m}+\sum_{i=x,y}\sum_{j=x,y,z}S_{ij}p^{i}%
\sigma ^{j}+V(\mathbf{r}),  \label{h_2D}
\end{equation}%
where $\mathbf{p}=(p^{x},p^{y})$ is the momentum operator in two dimensions, 
$m$ is the effective mass, and $\sigma ^{j}$'s with $j=x,y,z$ are the Pauli
matrices. The second term is the general form of SOC in the Cartesian
coordinate, where $S_{ij}$ is determined by SOCs linear in momentum such as
Rashba or Dresselhaus (001) terms. Here $V(\mathbf{r})$ represents the
potential confining electrons to the quasi-1D channel.

To obtain the effective Hamiltonian, we take the same approach as Refs.\ %
\onlinecite{Costa1981} and \onlinecite{Zhang2007}. Let $\mathbf{a}(q^{1})$
be the parametric equation of a planar curve where $q^{1}$ is the arc length
of the curve. The position of an electron in the plane can be written as 
\begin{equation*}
\mathbf{r}(q^{1},q^{2})=\mathbf{a}(q^{1})+q^{2}\hat{n}(q^{1}),
\end{equation*}%
where $\hat{n}(q^{1})$ is the unit normal vector of $\mathbf{a}(q^{1})$. 
$V(\mathbf{r})$ is of the form: 
\begin{equation*}
V({q^{2}})=%
\begin{cases}
0, & q^{2}=0 \\ 
\infty , & \text{else}%
\end{cases}%
.
\end{equation*}%
We are going to obtain the effective 1D Hamiltonian only depending on $q^{1}$%
. The steps are (i) to write the Hamiltonian in the curvilinear coordinates $%
q^{1}$ and $q^{2}$, (ii) to write an adequate transform of wave functions
and (iii) to take $q^{2}\rightarrow 0$. After these steps, the Hamiltonian
will be separated into two independent parts regarding $q^{1}$ and $q^{2}$,
respectively. Let $dl$ be an infinitesimal distance. We have 
\begin{equation*}
dl^{2}=g_{ij}dq^{i}dq^{j},
\end{equation*}%
where $g_{ij}$, the metric tensor, is defined by the inner product, $%
(\partial \mathbf{r}/\partial q^{i})\cdot (\partial \mathbf{r}/\partial
q^{j})$. We can use the coordinates $q$'s and the metric tensor $g^{ij}$ to
express the Laplacian: 
\begin{equation}
\nabla ^{2}\psi =\frac{1}{\sqrt{g}}\frac{\partial }{\partial q^{i}}\left[ 
\sqrt{g}g^{ij}\frac{\partial \psi }{\partial q^{j}}\right] ,
\label{laplacian}
\end{equation}%
where $g$ is the determinant of $g_{ij}$. Once we have $S_{ij}p^{i}\sigma
^{j}$ in the Cartesian coordinates, we can obtain the expression in the
coordinates $q$'s via the transform's laws of tensors: 
\begin{equation}
\begin{split}
p^{\prime i}& =\frac{\partial q^{\prime i}}{\partial q^{j}}p^{j}, \\
\sigma ^{\prime i}& =\frac{\partial q^{\prime i}}{\partial q^{j}}\sigma ^{j},
\\
S_{\mu \nu }^{^{\prime }}& =\frac{\partial q^{i}}{\partial q^{\prime \mu }}%
\frac{\partial q^{j}}{\partial q^{\prime \nu }}S_{ij},
\end{split}
\label{law_of_transform}
\end{equation}%
where the primed symbols denote those in the new coordinates. Using Eqs.\ %
\eqref{laplacian} and \eqref{law_of_transform}, we can write the Hamiltonian
of Eq.\ \eqref{h_2D} in the coordinates $q$'s:%
\begin{eqnarray}
\mathcal{H} &=&-\frac{\hbar ^{2}}{2m}\frac{1}{\sqrt{g}}\frac{\partial }{%
\partial q^{1}}\left( \sqrt{g}g^{11}\frac{\partial }{\partial q^{1}}\right) 
\notag \\
&&-i\hbar \sum_{j=1}^{3}S_{1j}\sigma ^{j}g^{11}\frac{\partial }{\partial
q^{1}}  \notag \\
&&-\frac{\hbar ^{2}}{2m}\left[ \frac{\partial ^{2}}{\partial {q^{2}}^{2}}+%
\frac{\partial }{\partial q^{2}}(\ln \sqrt{g})\frac{\partial }{\partial q^{2}%
}\right]  \notag \\
&&-i\hbar \sum_{j=1}^{3}S_{2j}\sigma ^{j}g^{22}\frac{\partial }{\partial
q^{2}}+V(q^{2}),  \label{h_curvlinear_conf}
\end{eqnarray}%
where for brevity the prime is neglected. We have used $g^{i2}=g^{2i}=\delta
_{2i}$ in Eq.\ \eqref{h_curvlinear_conf}. The first two terms and latter two
of Eq.\ \eqref{h_curvlinear_conf} are not independent since $g$ is a
function of $q^{1}$ and $q^{2}$. Given an eigenfunction $\Psi (q^{1},q^{2})$
of $\mathcal{H}$, we have $\mathcal{H}\Psi =E\Psi .$ Following Ref.\ %
\onlinecite{Costa1981}, we make the transform $\chi
(q^{1},q^{2})=f^{1/2}\Psi (q^{1},q^{2})$ with $f=\sqrt{g}=(1-\kappa q^{2})$,
where $\kappa $ is the curvature of $\mathbf{a}(q^{1})$. After the
transform, we obtain%
\begin{eqnarray}
\mathcal{H}\chi &=&\sqrt{f}\left[ -\frac{\hbar ^{2}}{2m}\frac{1}{\sqrt{g}}%
\frac{\partial }{\partial q^{1}}\left( \sqrt{g}g^{11}\frac{\partial }{%
\partial q^{1}}\frac{\chi }{\sqrt{f}}\right) \right]  \notag \\
&&-i\hbar \sum_{j=1}^{3}S_{1j}\sigma ^{j}g^{11}\frac{\partial \chi }{%
\partial q^{1}}  \notag \\
&&-\frac{\hbar ^{2}}{2m}\left\{ \frac{\partial ^{2}\chi }{\partial {q^{2}}%
^{2}}+\frac{1}{4f^{2}}\left[ \left( \frac{\partial f}{\partial q^{2}}\right)
^{2}\right] \chi \right\}  \notag \\
&&-i\hbar \sum_{j=1}^{3}S_{1j}\sigma ^{j}g^{22}\left( \frac{\partial }{%
\partial q^{2}}+\frac{1}{2f}\kappa \right) \chi +V(q^{2})\chi .
\label{h_trans}
\end{eqnarray}%
Taking $q^{2}\rightarrow 0$ except that in $V(q^{2})$, Eq.\ \eqref{h_trans}
becomes 
\begin{align}
\mathcal{H}\chi & =-\frac{\hbar ^{2}}{2m}\frac{\partial ^{2}\chi }{\partial {%
q^{1}}^{2}}-i\hbar \sum_{j=1}^{3}S_{1j}(q^{1},0)\sigma ^{j}\frac{\partial
\chi }{\partial q^{1}}  \notag \\
& -\frac{\hbar ^{2}}{2m}\left( \frac{\partial ^{2}\chi }{\partial {q^{2}}^{2}%
}+\frac{\kappa ^{2}\chi }{4}\right)  \notag \\
& -i\hbar \sum_{j=1}^{3}S_{2j}(q^{1},0)\sigma ^{j}\left( \frac{\partial \chi 
}{\partial q^{2}}+\frac{\kappa \chi }{2}\right) +V(q^{2})\chi .
\label{h_limit}
\end{align}%
%
%
%
%
%
%
%
%
%
%
%
%
%
%
%
%
%
%
%
%
%
%
%
%
%
%
%
%
%
%
%
%
%
%
%
%
%
%
%
%
%
%
%
%
%
%
%
%
%
%
%
%
%
%
%
%
%
%
%
%
%
%
%
%
%
%
%
%
%
%
%
%
%
%
%
%
Renaming $q^{1}$ as $s$ and deleting the terms dependent on $q^{2}$ in Eq.\ %
\eqref{h_limit}, we obtain the 1D effective Hamiltonian%
\begin{eqnarray}
\mathcal{H}_{1D} &=&-\frac{\hbar ^{2}}{2m}\frac{\partial ^{2}}{\partial s^{2}%
}-\frac{\hbar ^{2}\kappa ^{2}}{8m}  \notag \\
&&-i\hbar \mathbf{S}_{\parallel }(s)\cdot \vec{\sigma}\frac{\partial }{%
\partial s}-\frac{i\hbar \kappa }{2}\mathbf{S}_{\perp }(s)\cdot \vec{\sigma},
\label{H1D general}
\end{eqnarray}%
where $s$ denotes the arc length of the wire, and $\mathbf{S}_{\parallel }$
and $\mathbf{S}_{\perp }$ are defined by 
\begin{align}
\mathbf{S}_{\parallel }\cdot \vec{\sigma}&
=\sum_{j=1}^{3}S_{1j}(q^{1},0)\sigma ^{j},  \label{S_parallel} \\
\mathbf{S}_{\perp }\cdot \vec{\sigma}& =\sum_{j=1}^{3}S_{2j}(q^{1},0)\sigma
^{j}.  \label{S_perp}
\end{align}%
The second term in Eq.\ \eqref{H1D general} is the curvature-induced
geometric potential, which was first introduced by da Costa\cite{Costa1981}
and was previously neglected in the literature of mesoscopic ring transport.%
\cite%
{Stern1992,Aronov1993,Meijer2002,Splettstoesser2003,Frustaglia2004,Molnar2004,Souma2004,Wang2005}
We will later come back to investigate the role played by this geometric
potential term.

\subsubsection{1D arc with SOC: Continuous form}

Below we consider the Rashba SOC in an arc. In a two-dimensional electron
gas (2DEG), the intensively discussed Rashba SOC\cite{Bychkov1984} reads

\begin{equation}
\mathcal{H}_{R}^{2D}=\frac{\alpha }{\hbar }(p^{y}\sigma ^{x}-p^{x}\sigma
^{y}),  \label{HR 2D}
\end{equation}%
where $\alpha $ is the Rashba coupling parameter. In this case, $S_{ij}$ are%
\begin{equation}
\begin{split}
S_{xx}& =0, \\
S_{xy}& =\frac{-\alpha }{\hbar }, \\
S_{yx}& =\frac{\alpha }{\hbar }, \\
S_{yy}& =0.
\end{split}
\label{S}
\end{equation}%
The parametric equation for an arc can be written as $\mathbf{R}=r\hat{\rho}%
, $ where $r$ is the radius of the ring. The position of an electron is
written as $\mathbf{R}_{e}=(r-q^{2})\hat{\rho}.$ Here, $q^{1}$ is $r\phi $.
Using the transform law of tensors, we obtain $S_{ij}^{\prime }$ in the $%
q^{i}$ coordinates from Eqs.\ \eqref{S}, and thus%
\begin{equation}
\begin{split}
S_{11}& =0, \\
S_{12}& =\frac{-\alpha }{\hbar }, \\
S_{21}& =\frac{\alpha }{\hbar }, \\
S_{22}& =0.
\end{split}
\label{S'}
\end{equation}%
Using Eqs.\ \eqref{H1D general}, \eqref{S_parallel}, \eqref{S_perp}, and %
\eqref{S'}, we obtain the confined $\mathcal{H}_{R}$ in the polar
coordinates,%
\begin{equation*}
\mathcal{H}_{R}=-i\alpha \left( -\sigma ^{2}\frac{\partial }{r\partial \phi }%
+\frac{\sigma ^{1}}{2r}\right) ,
\end{equation*}%
where $\sigma ^{1}$ and $\sigma ^{2}$ are defined by Eqs.\ %
\eqref{law_of_transform}, or in the Cartesian coordinates,%
\begin{eqnarray}
\mathcal{H}_{R} &=&-\frac{i\alpha }{r}\left( \cos \phi \sigma ^{x}+\sin \phi
\sigma ^{y}\right) \frac{\partial }{\partial \phi }  \notag \\
&&-\frac{i\alpha }{2r}\left( \cos \phi \sigma ^{y}-\sin \phi \sigma
^{x}\right) ,  \label{HR for ring}
\end{eqnarray}%
which is in agreement with the terms given by Meijer \textit{et al.}\cite%
{Meijer2002}

The linear Dresselhaus (001) term, in a 2DEG expressed as\cite%
{Dresselhaus1955,Dyakonov1971}%
\begin{equation}
\mathcal{H}_{D}^{2D}=\frac{\beta }{\hbar }(p^{y}\sigma ^{y}-p^{x}\sigma
^{x}),  \label{HD 2D}
\end{equation}%
can be derived similarly for the arc, but can be written even more
conveniently by replacing from $\mathcal{H}_{R}$ with $\alpha \rightarrow
\beta ,\sigma ^{x}\rightarrow \sigma ^{y},\sigma ^{y}\rightarrow \sigma ^{x}$%
:%
\begin{eqnarray}
\mathcal{H}_{D} &=&-\frac{i\beta }{r}\left( \cos \phi \sigma ^{y}+\sin \phi
\sigma ^{x}\right) \frac{\partial }{\partial \phi }  \notag \\
&&-\frac{i\beta }{2r}\left( \cos \phi \sigma ^{x}-\sin \phi \sigma
^{y}\right) .  \label{HD for ring}
\end{eqnarray}%
Thus the 1D Hamiltonian in an arc in the presence of Rashba and linear
Dresselhaus (001) SOCs reads%
\begin{equation}
\mathcal{H}=-\frac{\hbar ^{2}}{2m}\frac{\partial ^{2}}{\partial s^{2}}-\frac{%
\hbar ^{2}\kappa ^{2}}{8m}+\mathcal{H}_{so},  \label{H}
\end{equation}%
where $\mathcal{H}_{so}$ is given in Eqs.\ \eqref{HR for ring} and 
\eqref{HD
for ring}.

\subsubsection{1D arc with SOC: Tight-binding form\label{sec H}}

Previously Souma and Nikoli\'{c} derived the tight-binding Hamiltonian for
two-dimensional rings in the presence of Rashba SOC.\cite{Souma2004}
Following their construction, here we take the 1D limit, add the previously
absent geometric potential term [second term in Eq.\ \eqref{H1D general} or %
\eqref{H}] and the linear Dresselhaus (001) term, to obtain%
\begin{eqnarray}
H &=&\left( U+2t_{0}+U_{g}\right) \sigma ^{0}\sum_{n}c_{n}^{\dag }c_{n} 
\notag \\
&&+\sum_{n}\left( \mathbf{t}_{n\leftarrow n+1}c_{n}^{\dag }c_{n+1}+\text{H.c.%
}\right) ,  \label{H ring}
\end{eqnarray}%
with the hopping matrix%
\begin{eqnarray}
\mathbf{t}_{n\leftarrow n+1} &=&-t_{0}\sigma ^{0}+i[\cos \phi _{n,n+1}\left(
t_{R}\sigma ^{x}+t_{D}\sigma ^{y}\right)  \notag \\
&&+\sin \phi _{n,n+1}\left( t_{R}\sigma ^{y}+t_{D}\sigma ^{x}\right) ].
\label{tnm}
\end{eqnarray}%
Here $t_{0}=\hbar ^{2}/2ma^{2},$ $a$ being the lattice grid spacing, is the
kinetic hopping parameter, $\sigma ^{0}$ is the $2\times 2$ identity matrix, 
$t_{R}=\alpha /2a,t_{D}=\beta /2a$ are the Rashba and Dresselhaus hopping
parameters, respectively, and $\phi _{n,n+1}=\left( \phi _{n}+\phi
_{n+1}\right) /2$ is the average azimuthal angle between site $n$ and site $%
n+1$ ($\phi _{n+1}>\phi _{n}$; see Ref.\ \onlinecite{Souma2004}). In the
on-site potential term in Eq.\ \eqref{H ring}, $U+2t_{0}$ responsible for
energy band offset corresponds to the atomic orbital energy in the language
of empirical tight-binding band calculation. In general $U$ can also take
into account other local potentials, but here for convenience we will put $U$
to zero. The additional term $U_{g}$ is the geometric potential and can be
reexpressed in terms of $t_{0}$ as%
\begin{equation}
U_{g}=-\frac{\hbar ^{2}\kappa ^{2}}{8m}=-\left( \frac{\pi }{2N_{r}}\right)
^{2}t_{0},  \label{Ug}
\end{equation}%
where relations $\kappa =1/r$ and $N_{r}a=\pi r$ are used. Note that the $%
U_{g}$ term will be later considered only in the LKF, but not other quantum
mechanical approaches.

\subsection{Spin transport formalisms}

Below we briefly review a set of different formalisms to be used to study
the charge and spin transport in the U-channel. We will first introduce the
tight-binding-based LKF (Sec.\ \ref{sec lkf}), for which the U-channel is
precisely described by Fig.\ \ref{fig1}(a). That is, a ferromagnetic lead is
attached to the left end of the U-channel, while the right lead is made of
normal metal; a bias potential difference is applied between the leads so
that electrical spin injections from the left lead are theoretically
simulated. Contrary to the sophisticated LKF, the quantum-mechanics-based
translation (Sec.\ \ref{sec translation}), as well as its approximating form
the SVF (Sec.\ \ref{sec svf}), and the spin-orbit gauge method (Sec.\ \ref%
{sec sog}) are schematically described by Fig.\ \ref{fig1}(b). That is, we
simply assume an ideal spin injected at the left end of the channel, drag
the spin through a U-shaped path using either a space translation operator
or a more elegant spin-orbit gauge operator, and then see how the spin
direction changes along the path.

\subsubsection{Landauer--Keldysh formalism\label{sec lkf}}

The key role in the LKF is played by the lesser Green's function, which
requires (i) a tight-binding Hamiltonian and (ii) lead self-energy. For (i),
the Hamiltonian matrix for the half-ring part has been introduced in Sec.\ %
\ref{sec H}. That for the arm parts (regions I and III) can be
straightforwardly constructed from the first-quantized Hamiltonians of Eqs.\ %
\eqref{HR 2D} and \eqref{HD 2D} and will not be repeated here. The size of
the full Hamiltonian matrix $\left[ H\right] $ amounts to $N\times N$, where 
$N=2N_{w}+N_{r}$ is the total number of sites. Each matrix element is a $%
2\times 2$ matrix because we are considering spin--%
$\frac12$
systems. For (ii), we consider semi-infinite discrete leads and summarize
the self-energy expression as follows.

Consider a ferromagnetic semi-infinite chain with uniform magnetization
pointing along $\mathbf{e}_{M}=(\sin \theta _{M}\cos \phi _{M},\sin \theta
_{M}$ $\sin \phi _{M},\cos \theta _{M})$. Extending the nonmagnetic and
continuous case from Ref.\ \onlinecite{Datta1995} to a ferromagnetic and
discrete one, we obtain%
\begin{equation}
\begin{split}
\Sigma _{M}\left( E\right) & =t_{c}^{2}g_{M}^{R}\left( E\right) , \\
g_{M}^{R}\left( E\right) & =\sum_{\sigma =\pm }g^{R}\left( E-\sigma
t_{M}\right) |\sigma ;\mathbf{e}_{M}\rangle \langle \sigma ;\mathbf{e}_{M}|,
\end{split}
\label{self-energy}
\end{equation}%
where $t_{c}$ is the coupling strength between the lead and the central
transport channel (and will be set equal to $t_{0}$), $t_{M}$ is the Zeeman
splitting energy, the eigenkets are\cite{Sakurai1994}%
\begin{align*}
|\sigma & =+;\mathbf{e}_{M}\rangle =\left( 
\begin{array}{c}
e^{-i\phi _{M}}\cos \dfrac{\theta _{M}}{2} \\ 
\sin \dfrac{\theta _{M}}{2}%
\end{array}%
\right) , \\
|\sigma & =-;\mathbf{e}_{M}\rangle =\left( 
\begin{array}{c}
e^{-i\phi _{M}}\sin \dfrac{\theta _{M}}{2} \\ 
-\cos \dfrac{\theta _{M}}{2}%
\end{array}%
\right) ,
\end{align*}%
and the retarded surface Green's function reads 
\begin{align*}
g^{R}\left( E\right) & =\frac{1}{2t_{d}}%
\begin{cases}
\Delta -i\sqrt{4-\Delta ^{2}}, & \left\vert \Delta \right\vert \leq 2 \\ 
\Delta -\limfunc{sgn}\Delta \sqrt{\Delta ^{2}-4}, & \left\vert \Delta
\right\vert >2%
\end{cases}%
, \\
\Delta & =\frac{E-\left( V+2t_{d}\right) }{t_{d}},
\end{align*}%
where $t_{d}$ is the kinetic hopping parameter in the lead and will be again
set equal to $t_{0}$ in the later computation. The self-energy function,
Eq.\ \eqref{self-energy}, is the only nonvanishing matrix element of the
full self-energy matrices: $\left[ \Sigma _{L}(E)\right] _{11}$ and $\left[
\Sigma _{R}(E)\right] _{NN}$. For our U-channel here, we will consider for
the left lead $t_{M}=0.1t_{0}$ to inject spins while for the right lead $%
t_{M}=0$ to let the spins outflow freely.

With both the tight-binding Hamiltonian and lead self-energy matrix
constructed, one can construct the space-resolved retarded Green's function
matrix%
\begin{equation*}
\left[ G^{R}\left( E\right) \right] =\left\{ E\left[ I\right] -\left[ H%
\right] -\left[ \Sigma _{L}\left( E\right) \right] -\left[ \Sigma _{R}\left(
E\right) \right] \right\} ^{-1},
\end{equation*}%
where $\left[ I\right] $ is the $2N\times 2N$ identity matrix, $\left[ H%
\right] $ is the space-resolved tight-binding Hamiltonian matrix for the
U-channel, and $\left[ \Sigma _{L/R}\left( E\right) \right] $ is the
self-energy matrix of the left/right lead. The lesser Green's function
matrix is then obtained via the kinetic equation%
\begin{equation}
\left[ G^{<}\left( E\right) \right] =\left[ G^{R}\left( E\right) \right] %
\left[ \Sigma ^{<}\left( E\right) \right] \left[ G^{A}\left( E\right) \right]
,  \label{lesser G}
\end{equation}%
where $\left[ G^{A}\left( E\right) \right] $ is the advanced Green's
function matrix obtained by the Hermitian conjugate of $\left[ G^{R}\left(
E\right) \right] $ and the lesser self-energy matrix is given by%
\begin{eqnarray*}
\left[ \Sigma ^{<}\left( E\right) \right] &=&-\sum_{p=L,R}\left\{ \left[
\Sigma _{p}\left( E-eV_{p}\right) \right] -\left[ \Sigma _{p}\left(
E-eV_{p}\right) \right] ^{\dag }\right\} \\
&&\times f_{0}\left( E-eV_{p}\right) ,
\end{eqnarray*}%
where $f_{0}$ is the Fermi function and $eV_{p}$ is the electric potential
energy applied on lead $p$. In our numerical computation we will put $%
eV_{L}=+eV_{0}/2$ and $eV_{R}=-eV_{0}/2$ for a potential energy difference
of $eV_{0}$, a bias parameter that is taken as positive (while the electron
charge $e=-\left\vert e\right\vert $ is negative), so that the electrons are
injected from the left lead. In addition, we will consider a
zero-temperature limit so that the Fermi function becomes step-like and will
strictly cut the energy integration range [see Eqs.\ \eqref{charge density}
and \eqref{spin
density} below].

Desired physical quantities can then be extracted from the lesser Green's
function, Eq.\ \eqref{lesser G}, through properly defined expressions.\cite%
{Nikolic2006} In this paper our main interest lies in the local charge
density,%
\begin{equation}
e\langle N_{n}\rangle =\frac{e}{2\pi i}%
\int_{E_{F}-eV_{0}/2}^{E_{F}+eV_{0}/2}dE\func{Tr}_{s}\left[ G^{<}\left(
E\right) \right] _{nn},  \label{charge density}
\end{equation}%
and the local spin density,%
\begin{equation}
\begin{split}
\langle S_{n}^{i}\rangle =& \frac{\hbar /2}{2\pi i}%
\int_{E_{F}-eV_{0}/2}^{E_{F}+eV_{0}/2}dE\func{Tr}_{s}\left\{ \sigma ^{i}%
\left[ G^{<}\left( E\right) \right] _{nn}\right\} , \\
i=& x,y,z
\end{split}
\label{spin density}
\end{equation}%
where $E_{F}$ is the Fermi level that will be set to $0.2t_{0}$ above the
band bottom, $\left[ G^{<}\left( E\right) \right] _{nn}$ is the $n$th
diagonal matrix element of the entire $\left[ G^{<}\left( E\right) \right] $
matrix and is a $2\times 2$ matrix, and $\func{Tr}_{s}$ is the trace done
with respect to spin. The subscript on the left-hand sides of both Eqs.\ %
\eqref{charge density} and \eqref{spin density} stand for the $n$th site of
the U-channel.

\subsubsection{Quantum mechanical translation method\label{sec translation}}

In the following we briefly review an earlier work done by some of us,\cite%
{Liu2006b} a theoretical method based on quantum mechanics to analyze spin
precession along an arbitrary path.

An electron spin injected at $\mathbf{r}_{0}$ is described by a state ket $|%
\mathbf{s}_{0};\mathbf{r}_{0}\rangle ,$ where $\mathbf{s}_{0}$ labels the
spin orientation, and is later evolved to another state ket $|\mathbf{s};%
\mathbf{r}\rangle $ at position $\mathbf{r}$, through the translation
operator $\mathcal{T}\left( \mathbf{p}\right) =\exp \left[ i\mathbf{p}/\hbar
\cdot \left( \mathbf{r}-\mathbf{r}_{0}\right) \right] $, i.e., $|\mathbf{s};%
\mathbf{r}\rangle =\mathcal{T}\left( \mathbf{p}\right) |\mathbf{s}_{0};%
\mathbf{r}_{0}\rangle $. In two-dimensional boundless systems with Rashba
and linear Dresselhaus (001) SOCs [Eqs.\ \eqref{HR 2D} and \eqref{HD 2D}],
the eigenstates $|\pm ;\phi _{k}\rangle $ are well known (see, for example,
also Ref.\ \onlinecite{Liu2006b}) and can serve as a convenient basis to
expand the spin state ket; $\phi _{k}$ is the propagation angle of wave
vectors $\mathbf{k}_{\pm }$. Hence, expanding $|\mathbf{s}_{0};\mathbf{r}%
_{0}\rangle $ in terms of $|\pm ;\phi _{k}\rangle $, we can proceed by using 
$f\left( \mathbf{p}\right) |\pm ;\phi _{k}\rangle =f\left( \hbar \mathbf{k}%
_{\pm }\right) |\pm ;\phi _{k}\rangle $:%
\begin{eqnarray}
|\mathbf{s};\mathbf{r}\rangle &=&\mathcal{T}\left( \mathbf{p}\right) |%
\mathbf{s}_{0};\mathbf{r}_{0}\rangle  \notag \\
&=&e^{i\mathbf{k}_{\pm }\cdot \left( \mathbf{r}-\mathbf{r}_{0}\right)
}\sum_{\sigma =\pm }|\sigma ;\phi _{k}\rangle \langle \sigma ;\phi _{k}|%
\mathbf{s}_{0};\mathbf{r}_{0}\rangle  \notag \\
&=&e^{i\bar{k}\Delta r}\sum_{\sigma =\pm }e^{i\sigma \Delta \theta
/2}|\sigma ;\phi _{k}\rangle \langle \sigma ;\phi _{k}|\mathbf{s}_{0};%
\mathbf{r}_{0}\rangle ,  \label{state ket}
\end{eqnarray}%
with $\bar{k}=(k_{+}+k_{-})/2,\Delta \theta =\Delta k\Delta
r=(k_{+}-k_{-})\Delta r,$ and $\Delta r=\left\vert \mathbf{r}-\mathbf{r}%
_{0}\right\vert $. The global phase involving $\bar{k}$ will be canceled in
calculating the expectation value while the phase difference involving $%
\Delta k=-2m\zeta /\hbar ^{2}$ with $\zeta $ given later in Eq.\ \eqref{zeta}
plays a key role in spin precession. For successive nearest-neighbor
hoppings in Fig.\ \ref{fig1}(a), we simply apply Eq.\ \eqref{state ket} for
every step and then calculate the expectation value for Pauli matrices to
obtain the spin direction on each site, $\langle \mathbf{S}\rangle =(\hbar
/2)\langle \vec{\sigma}\rangle =(\hbar /2)\langle \mathbf{s};\mathbf{r}%
|(\sigma _{x},\sigma _{y},\sigma _{z})|\mathbf{s};\mathbf{r}\rangle ,$
starting with the assumed injected spin at the first site in contact with
the left lead.

\subsubsection{Spin vector formula\label{sec svf}}

A further approximating step done in Ref.\ \onlinecite{Liu2006b} (see the
Appendix therein) was to take the continuous limit, so that each section
approaches to infinitesimal. After successive infinitesimal translations
from injection point $\mathbf{r}_{0}$ to a certain desired position $\mathbf{%
r}$, the spinor overlaps carried by the final state ket were approximated as%
\begin{eqnarray}
&&\langle \sigma _{1}|\sigma _{0}\rangle \langle \sigma _{2}|\sigma
_{1}\rangle \cdots \langle \sigma _{j+1}|\sigma _{j}\rangle \cdots \langle
\sigma _{N}|\sigma _{N-1}\rangle   \notag \\
&\approx &\langle \sigma _{1}|\sigma _{0}\rangle \delta _{\sigma _{2}\sigma
_{1}}\cdots \delta _{\sigma _{j+1}\sigma _{j}}\cdots \delta _{\sigma
_{N}\sigma _{N-1}},  \label{overlap}
\end{eqnarray}%
where $|\sigma _{0}\rangle =|\mathbf{s}_{0};\mathbf{r}_{0}\rangle $ is the
input, $|\sigma _{j}\rangle $ is the shorthand for $|\sigma _{j};\phi
_{k}^{j}\rangle $, $\phi _{k}^{j}$ being the propagation angle of the $j$th
section, and $N\rightarrow \infty $ is the number of infinitesimal straight
translations from $\mathbf{r}_{0}$ to $\mathbf{r}$. A closed form of the
state ket generalized from Eq.\ \eqref{state
ket} can thus be obtained. Using the generalized state ket one obtains the
SVF,%
\begin{widetext}%
\begin{equation}
\langle \mathbf{S}\rangle =\frac{\hbar }{2}\left( 
\begin{array}{c}
-\cos \theta _{M}\cos \varphi _{k}\sin \Delta \Theta +\sin \theta _{M}\left[
\cos \left( \varphi _{k}-\varphi _{k}^{0}+\phi _{M}\right) \cos ^{2}\dfrac{%
\Delta \Theta }{2}-\cos \left( \varphi _{k}+\varphi _{k}^{0}-\phi
_{M}\right) \sin ^{2}\dfrac{\Delta \Theta }{2}\right]  \\ 
-\cos \theta _{M}\sin \varphi _{k}\sin \Delta \Theta +\sin \theta _{M}\left[
\sin \left( \varphi _{k}-\varphi _{k}^{0}+\phi _{M}\right) \cos ^{2}\dfrac{%
\Delta \Theta }{2}-\sin \left( \varphi _{k}+\varphi _{k}^{0}-\phi
_{M}\right) \sin ^{2}\dfrac{\Delta \Theta }{2}\right]  \\ 
\cos \theta _{M}\cos \Delta \Theta +\sin \theta _{M}\cos \left( \varphi
_{k}^{0}-\phi _{M}\right) \sin \Delta \Theta 
\end{array}%
\right) ,  \label{SVF}
\end{equation}%
\end{widetext}%
with 
\begin{equation}
\varphi _{k}=\arg [(\alpha \cos \phi _{k}-\beta \sin \phi _{k})+i(\alpha
\sin \phi _{k}-\beta \cos \phi _{k})],  \label{vphi}
\end{equation}%
\begin{equation}
\Delta \Theta =\frac{2m^{\star }}{\hbar ^{2}}\int_{C}\zeta ds,
\label{Dtheta}
\end{equation}
\begin{equation}
\zeta =\sqrt{\alpha ^{2}+\beta ^{2}-2\alpha \beta \sin 2\phi _{k}\left(
s\right) }.  \label{zeta}
\end{equation}%
The angle $\varphi _{k}^{0}$ in Eq.\ \eqref{SVF} stands for $\varphi
_{k}\left( \phi _{k}^{0}\right) ,$ where $\phi _{k}^{0}$ is the propagation
direction of the input $|\mathbf{s}_{0};\mathbf{r}_{0}\rangle $.

For the present U-channel, the transport direction as a function of position
coordinate $s$ can be written as%
\begin{equation}
\phi _{k}\left( s\right) =%
\begin{cases}
\pi /2, & s\in \left[ 0,L\right]  \\ 
\pi /2-\pi \frac{s-L}{\pi r}, & s\in \left[ L,L+\pi r\right]  \\ 
-\pi /2, & s\in \lbrack L+\pi r,2L+\pi r]%
\end{cases}%
,  \label{phi}
\end{equation}%
$L$ being the length of each arm; $s$ runs from $0$ to $2L+\pi r$. In the
following we give two concrete examples to show the convenience of Eq.\ %
\eqref{SVF}, one for the pure Rashba case and the other for the pure
Dresselhaus, both with $S^{x}$ spin injection: $(\theta _{M},\phi _{M})=(\pi
/2,0)$.

In the presence of only the Rashba SOC, we have from Eq.\ \eqref{vphi} $%
\varphi _{k}^{0}=\phi _{k}\left( s=0\right) =\pi /2,$ $\varphi _{k}=\phi
_{k}\left( s\right) $ and from Eqs.\ \eqref{Dtheta} and \eqref{zeta} $\Delta
\Theta =2(t_{R}/t_{0})(s/a)$. Putting these together with $(\theta _{M},\phi
_{M})=(\pi /2,0)$ into Eq.\ \eqref{SVF} we have%
\begin{equation}
\left. \langle \mathbf{S}\rangle \right\vert _{R,S^{x}\text{ inj}}=\frac{%
\hbar }{2}\left( 
\begin{array}{c}
\sin \phi _{k} \\ 
-\cos \phi _{k} \\ 
0%
\end{array}%
\right) .  \label{SVF R}
\end{equation}%
In the presence of only the linear Dresselhaus (001) SOC, we have $\varphi
_{k}^{0}=-\pi ,\varphi _{k}=-\phi _{k}(s)-\pi /2,$ and $\Delta \Theta
=2(t_{D}/t_{0})(s/a)$. Equation \eqref{SVF} then reduces to%
\begin{equation}
\left. \langle \mathbf{S}\rangle \right\vert _{D,S^{x}\text{ inj}}=\frac{%
\hbar }{2}\left( 
\begin{array}{c}
\cos \Delta \Theta \sin \phi _{k} \\ 
\cos \Delta \Theta \cos \phi _{k} \\ 
-\sin \Delta \Theta 
\end{array}%
\right) .  \label{SVF D}
\end{equation}

Despite the elegant description of these SVFs, a crucial approximation of
the spinor overlaps that has been made in Eq.\ \eqref{overlap} deserves a
further discussion before we move on. Take one pair of the overlap, say,
between $j$th and $(j+1)$th. Recall the eigenspinors in the presence of both
Rashba and linear Dresselhaus (001) SOCs,\cite{Liu2006b}%
\begin{equation}
|\sigma _{j}\rangle =\frac{1}{\sqrt{2}}\left( 
\begin{array}{c}
ie^{-i\varphi _{k}^{j}} \\ 
\sigma _{j}%
\end{array}%
\right) ,  \label{eigenspinor}
\end{equation}%
where $\varphi _{k}^{j}=\varphi _{k}(\phi _{k}^{j})$ is given in Eq.\ %
\eqref{vphi}. When the two sections point along the same direction, i.e., $%
\phi _{k}^{j}=\phi _{k}^{j+1}$, the orthogonality becomes exact: $\langle
\sigma _{j+1}|\sigma _{j}\rangle =\delta _{\sigma _{j+1}\sigma _{j}}$,
regardless of the type of the SOCs in the straight 1D structure. Otherwise,
the orthogonal approximation always contains an error. For the pure Rashba
case, the overlap using Eq.\ \eqref{eigenspinor} up to first order in $%
\Delta \phi _{k}$ reads $\langle \sigma _{j+1}|\sigma _{j}\rangle
=(e^{i(\phi _{k}^{j+1}-\phi _{k}^{j})}+\sigma _{j+1}\sigma _{j})/2=\delta
_{\sigma _{j+1}\sigma _{j}}+i\Delta \phi _{k}/2+\cdots ,$ which indicates
that the major error term accumulating upon \textquotedblleft
turning\textquotedblright\ along the curved 1D structure is proportional to
the change of the angle $\Delta \phi _{k}=\phi _{k}^{j+1}-\phi _{k}^{j}$ and
is therefore still moderate. In the presence of only the linear Dresselhaus
term, the situation is similar. In the presence of both SOC terms, however,
the error accumulated becomes drastic, which we will show numerically later.

\subsubsection{Spin-orbit gauge method\label{sec sog}}

The spin propagator can be obtained with the help of a spin-orbit gauge.\cite%
{Chen2008a} Noting that the highest order in momentum $\mathbf{P}=\mathbf{p}%
\sigma ^{0}$ in the Hamiltonian of a 2DEG with Rashba and Dresselhaus SOCs
(both linear in $\mathbf{p}$) is \emph{quadratic}, one can define the
spin-orbit gauge,%
\begin{equation}
\mathbf{A}^{\text{SO}}=\left( A_{x},A_{y}\right) \equiv \frac{mc}{e\hbar }%
\left( \alpha \sigma _{y}+\beta \sigma _{x},-\alpha \sigma _{x}-\beta \sigma
_{y}\right) ,  \label{eq: SOgauge}
\end{equation}%
to express the 2DEG Hamiltonian,%
\begin{eqnarray}
\mathcal{H}^{RD} &=&\frac{\mathbf{P}^{2}}{2m}+\mathcal{H}_{R}^{2D}+\mathcal{H%
}_{D}^{2D}  \notag \\
&=&\frac{1}{2m}\left( \mathbf{P-}\frac{e}{c}\mathbf{A}^{\text{SO}}\right)
^{2}-V_{b}\sigma ^{0}  \label{eq: RDgauge}
\end{eqnarray}%
with the constant background potential $V_{b}=(m/\hbar ^{2})\left( \alpha
^{2}+\beta ^{2}\right) $; recall that $\sigma ^{0}$ is the $2\times 2$
identity matrix. Consider now the transformation operator,%
\begin{equation}
\mathcal{U}^{\text{SO}}\left( \mathbf{r}\right) =\exp \left[ \frac{ie}{\hbar
c}\left( \mathbf{A}^{\text{SO}}\cdot \mathbf{r}\right) \right] ,
\label{eq: USOdef}
\end{equation}%
with the unitary property $\mathcal{U}^{\text{SO}}\left( \mathbf{r}\right) 
\mathcal{U}^{\text{SO}}\left( \mathbf{r}\right) ^{\dag }=\sigma ^{0}$
ensured by the Hermitian $\mathbf{A}^{\text{SO}}=\mathbf{A}^{\text{SO}\dag }$
from definition (\ref{eq: SOgauge}) and $\mathbf{r}$ being the position
vector of the electron displacement. Since the spin-orbit-interacting
Hamiltonian of Eq.\  (\ref{eq: RDgauge}) differs from the Hamiltonian of the
free electron gas (with a background potential $V_{b}$),%
\begin{equation}
\mathcal{H}^{\text{free}}=\frac{\mathbf{P}^{2}}{2m}-V_{b}\sigma ^{0}\text{,}
\label{eq: hBfree}
\end{equation}%
only by the gauge term $(e/c)\mathbf{A}^{\text{SO}}$ in $\mathbf{P}$, the
following transformation is therefore suggested:%
\begin{align}
& \mathcal{U}^{\text{SO}}\left( \mathbf{r}\right) \mathbf{P}\mathcal{U}^{%
\text{SO}}\left( \mathbf{r}\right) ^{\dag }=\mathbf{P+}\frac{ie}{\hbar c}%
\left[ \mathbf{A}^{\text{SO}}\cdot \mathbf{r,P}\right]   \notag \\
& +\frac{1}{2}\left( \frac{ie}{\hbar c}\right) ^{2}\left[ \mathbf{A}^{\text{%
SO}}\cdot \mathbf{r,}\left[ \mathbf{A}^{\text{SO}}\cdot \mathbf{r,P}\right] %
\right] +\cdots ,  \label{eq: Transform}
\end{align}%
with $[\mathbf{A}^{\text{SO}}\cdot \mathbf{r},\mathbf{P]}=i\hbar \mathbf{A}^{%
\text{SO}}$. Due to the noncommutability $[A_{x}^{\text{SO}},A_{y}^{\text{SO}%
}]\neq 0$, the terms containing higher orders of $\left\vert \mathbf{r}%
\right\vert $, in general, do not vanish, while in the small-displacement
limit $\left\vert \mathbf{r}\right\vert \approx 0$ in which one has $e\left(
c\hbar \right) ^{-1}\mathbf{A}^{\text{SO}}\cdot \mathbf{r}\ll 1$, Eq.\ %
\eqref{eq: Transform} reduces to $\mathcal{U}^{\text{SO}}\left( \mathbf{r}%
\right) \mathbf{P}\mathcal{U}^{\text{SO}}\left( \mathbf{r}\right) ^{\dag }=%
\mathbf{P-}e/c\mathbf{A}^{\text{SO}}$, rendering the following
transformation, 
\begin{equation}
\mathcal{U}^{\text{SO}}\left( \mathbf{r}\right) \mathcal{H}^{\text{free}}%
\mathcal{U}^{\text{SO}}\left( \mathbf{r}\right) ^{\dag }=\mathcal{H}^{RD}%
\text{,}  \label{eq: hfree_hRD}
\end{equation}%
between the two systems, $\mathcal{H}^{RD}$ and $\mathcal{H}^{\text{free}}$.
Accordingly, when $\left\vert \mathbf{r}\right\vert \approx 0$ or $e\left(
c\hbar \right) ^{-1}\mathbf{A}^{\text{SO}}\cdot \mathbf{r}\ll 1$ is
satisfied, the free electron gas $\mathcal{H}^{\text{free}}$, Eq.\ %
\eqref{eq: hBfree}, and the spin-orbit-interacting electron gas $\mathcal{H}%
^{RD}$, Eq.\ \eqref{eq: RDgauge}, share the same eigenenergies $E_{\mathbf{k}%
}$. Their corresponding eigenfunctions, denoted by $\psi _{E_{\mathbf{k}}}(%
\mathbf{r)}\chi _{s}^{\text{free}}$ and $\Psi _{E_{\mathbf{k}}}(\mathbf{r)}%
\chi _{s}^{\text{SO}}$, respectively, differ from each other only by a phase
factor, the $2\times 2$ matrix $\mathcal{U}^{\text{SO}}\left( \mathbf{r}%
\right) $, namely, $\Psi _{E_{\mathbf{k}}}(\mathbf{r)}\chi _{s}^{\text{SO}}=$
$\psi _{E_{\mathbf{k}}}(\mathbf{r)}\mathcal{U}^{\text{SO}}\left( \mathbf{r}%
\right) \chi _{s}^{\text{free}}$. Here $\chi _{s}$ is the spin part of the
wave function. Moreover, any wave function is constructed by a superposition
of the eigenfunctions, so for any given wave function $\psi (\mathbf{r)}\chi
_{s}^{\text{free}}$ in $\mathcal{H}^{\text{free}}$, the corresponding wave
function in $\mathcal{H}^{RD}$ is $\psi (\mathbf{r)}\mathcal{U}^{\text{SO}%
}\left( \mathbf{r}\right) \chi _{s}^{\text{free}}$.

The correspondence, which originated from gauge transformation (\ref{eq:
hfree_hRD}), between $\mathcal{H}^{\text{free}}$ and $\mathcal{H}^{RD}$
systems, allows one to construct the spin propagator for $\mathcal{H}^{RD}$;
to elaborate on this, consider an injected electron in system $\mathcal{H}^{%
\text{free}}$ described by $\psi _{\text{inj}}(\mathbf{r)}\chi _{\text{inj}%
}=[\sum_{_{\mathbf{k}}}C_{\mathbf{k}}\psi _{E_{\mathbf{k}}}(\mathbf{r)]}\chi
_{\text{inj}}$ with the initial spin state $\chi _{\text{inj}}$ and the
weighting factor $C_{\mathbf{k}}$. Without any spin-dependent mechanisms,
this electron remains at spin state $\chi _{\text{inj}}$, while importing $%
\mathbf{A}^{\text{SO}}$ turns on $\mathcal{U}^{\text{SO}}\left( \mathbf{r}%
\right) $ so that the electron wave function in the spin-orbit-interacting
system $\mathcal{H}^{RD}$ can be expressed by the gauge transformation in
the form,%
\begin{equation}
\mathcal{U}^{\text{SO}}\left( \mathbf{r}\right) \psi _{\text{inj}}\left( 
\mathbf{r}\right) \chi _{\text{inj}}=\sum_{\mathbf{k}}C_{\mathbf{k}}\psi
_{E_{\mathbf{k}}}\left( \mathbf{r}\right) \mathcal{U}^{\text{SO}}\left( 
\mathbf{r}\right) \chi _{\text{inj}}\text{.}  \label{eq: SOinj}
\end{equation}%
As a result, the spin polarization of the electron in system $\mathcal{H}%
^{RD}$ varies spatially according to $\mathcal{U}^{\text{SO}}\left( \mathbf{r%
}\right) \chi _{\text{inj}}$, and thus $\mathcal{U}^{\text{SO}}\left( 
\mathbf{r}\right) $ can be viewed as a spin propagator.

For general applications based on the gauge transformation, assume an
electron moving along an arbitrarily curved trajectory denoted as path $c$
starting from one spatial point to the other. Divide this path into $N$
pieces. Label the divided pieces (paths) by path $1$, path $2$, $\ldots $,
path $N$, sequentially (i.e., path $i+1$ follows path $i$), and let $\mathbf{%
r}_{i}$ denote the position vector of the displacement for the $i$th path.
One can always choose large-enough $N$ to have $e\left( c\hbar \right) ^{-1}%
\mathbf{A}^{\text{SO}}\cdot \mathbf{r}_{i}\ll 1$ such that Eq.\ 
\eqref{eq:
USOdef} can be approximately interpreted as a propagator for each $\mathbf{r}%
_{i}$. The spin propagator along an arbitrary path $c$ then reads%
\begin{equation}
\mathcal{U}_{c}^{\text{SO}}\left( \mathbf{r}\right) =\mathcal{U}^{\text{SO}%
}\left( \mathbf{r}_{N}\right) \mathcal{U}^{\text{SO}}\left( \mathbf{r}%
_{N-1}\right) \cdots \mathcal{U}^{\text{SO}}\left( \mathbf{r}_{2}\right) 
\mathcal{U}^{\text{SO}}\left( \mathbf{r}_{1}\right) \text{,}
\label{eq: genprop}
\end{equation}%
which can be concisely written as%
\begin{equation}
\mathcal{U}_{c}^{\text{SO}}\left( \mathbf{r}\right) =\mathcal{P}\exp \left( 
\frac{ie}{\hbar c}\int_{c}\mathbf{A}^{\text{SO}}\cdot d\mathbf{r}\right) 
\text{,}  \label{eq: UcSOdef}
\end{equation}%
where $\mathcal{P}$ is the path-ordering operator that orders the operator $%
\mathcal{U}^{\text{SO}}\left( \mathbf{r}_{i}\right) $ with earlier passing
path $\mathbf{r}_{i}$ to the right of the later $\mathcal{U}^{\text{SO}%
}\left( \mathbf{r}_{i+1}\right) $ such that $\mathcal{P}\exp [(ie/\hbar
c)\int_{\text{path }i+1\leftarrow \text{path }i}\mathbf{A}^{\text{SO}}\cdot d%
\mathbf{r}]=\mathcal{U}^{\text{SO}}\left( \mathbf{r}_{i+1}\right) \mathcal{U}%
^{\text{SO}}\left( \mathbf{r}_{i}\right) $.

Obviously, if both the $i$th and $(i+1)$th paths form a straight line, then
one has%
\begin{equation}
\mathcal{U}^{\text{SO}}\left( \mathbf{r}_{i+1}+\mathbf{r}_{i}\right) =%
\mathcal{U}^{\text{SO}}\left( \mathbf{r}_{i+1}\right) \mathcal{U}^{\text{SO}%
}\left( \mathbf{r}_{i}\right)   \label{eq: propprop}
\end{equation}%
simply because only one dimension (component) of $\mathbf{A}^{\text{SO}}$
will be used, and thus the commutators that appeared in the higher-order
terms of Eq.\ \eqref{eq: Transform} vanish. In other words, if the electron
moves along a straight line, i.e., path $c$ is not curved, we have $\mathcal{%
U}_{c}^{\text{SO}}\left( \mathbf{r}\right) =\mathcal{U}^{\text{SO}}\left( 
\mathbf{r}\right) $, namely, Eq.\ \eqref{eq: UcSOdef} reduces to Eq.\ %
\eqref{eq: USOdef}.

To study the spin evolution through the U-channel in the \emph{continuous }%
limit, one can use $\mathcal{U}^{\text{SO}}\left( \mathbf{r}\right) $ for
parts I and III and Eq.\ \eqref{eq: genprop} for part II. To be consistent
with the \emph{discrete} tight-binding model shown in Fig.\ \ref{fig1}(a)
adopted in the LKF, however, we will successively apply $\mathcal{U}^{\text{%
SO}}\left( \mathbf{r}\right) $ for each nearest-neighbor hopping.

\section{Transport analysis\label{sec transport analysis}}

Having reviewed the theoretical formalisms, we are now in a position to
carry out our three goals of this paper. In Sec.\ \ref{sec QMvsQT}, we
compare the spin precession patterns calculated by quantum mechanical
approaches with those by the LKF, or nowadays generally termed quantum
transport. Meanwhile, we will examine the role played by the
curvature-induced geometric potential based on the LKF. We proceed in Sec.\ %
\ref{sec spin flip} with a detailed discussion for adiabatic and
nonadiabatic transport regimes and connect the present paper with previous
ones. In Sec.\ \ref{sec charge and spin transport} we discuss the
anisotropic charge transport due to the interplay between the Rashba and
Dresselhaus SOCs, and spin precession in special cases, which is equivalent
to numerically testing the eigenstates of Rashba rings, Dresselhaus rings,
and the persistent spin-helix state.

\subsection{Quantum mechanical approaches vs quantum transport\label{sec
QMvsQT}}

\subsubsection{Weak geometric potential\label{sec weak Ug}}

\begin{figure*}[t]
\centering\includegraphics[width=\textwidth]{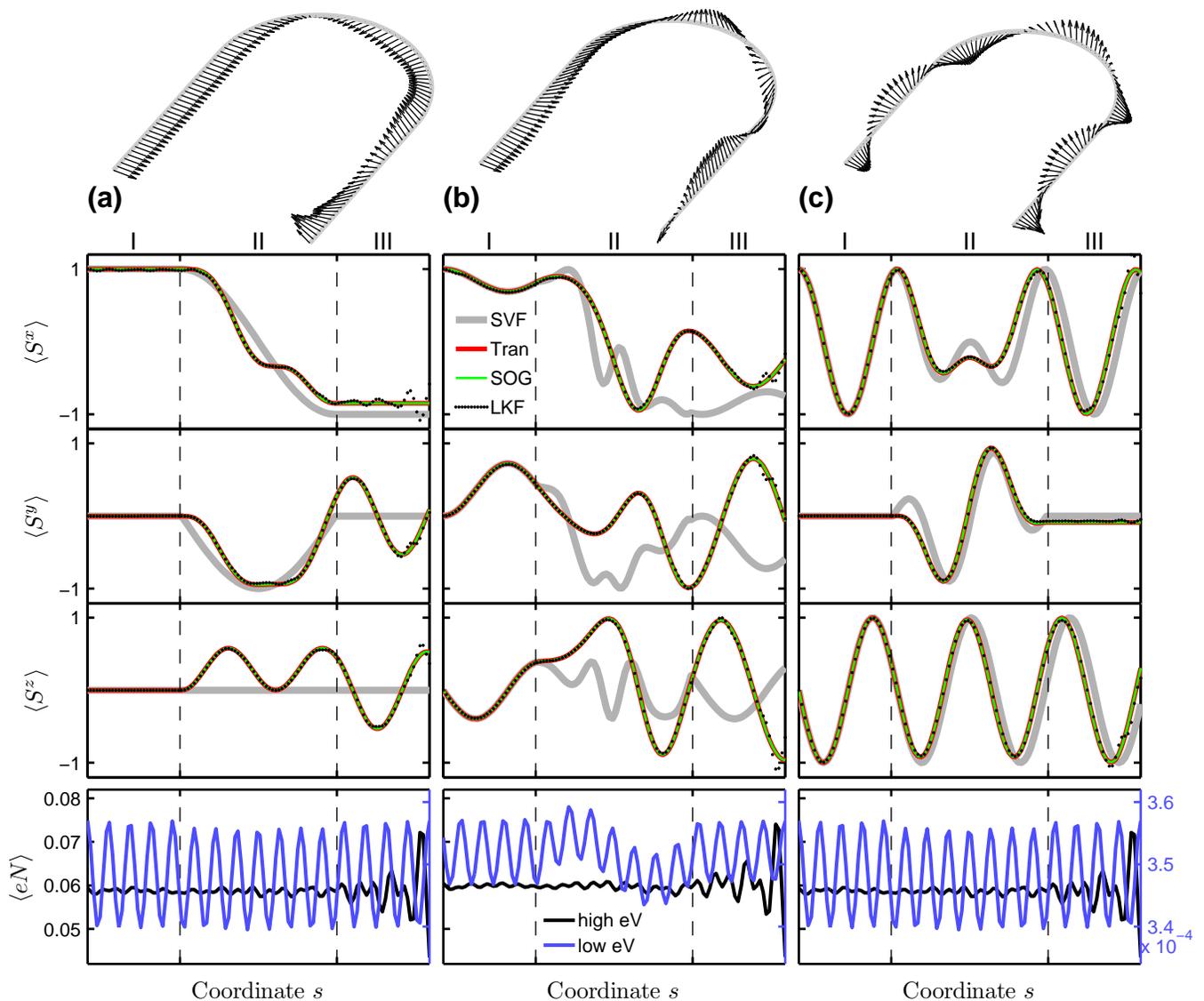}
\caption{(Color online) Spatially resolved spin components (in units of $%
\hbar /2$) calculated by the LKF, translation method (Tran), spin-orbit
gauge (SOG) method, and SVF in a U-channel with $(N_{w},N_{r})=(30,50)$
under a strong bias of $eV_{0}=0.4t_{0}$ and SOC strengths of (a) $%
(t_{R},t_{D})=(0.1,0)t_{0}$, (b) $(t_{R},t_{D})=(0.1,0.03)t_{0}$, and (c) $%
(t_{R},t_{D})=(0,0.1)t_{0}.$ The local charge densities obtained by the LKF
for a high bias of $eV_{0}=0.4t_{0}$ and low bias of $eV_{0}=10^{-3}t_{0}$
are shown in the bottom panels. Spin vectors in the top subplots are based
on the high-bias LKF results.}
\label{fig2}
\end{figure*}

Recall the geometric potential $U_{g}$ expressed in terms of $t_{0}$ in Eq.\ %
\eqref{Ug}. From the tight-binding Hamiltonian of Eq.\  \eqref{H ring}, one
can see that whether $U_{g}$ is sensible by the electrons depends on its
competition with the energy band width $2t_{0}$. Here we begin with a
U-channel with $N_{r}=50,$ yielding $\left\vert U_{g}\right\vert \approx
2.47\times 10^{-4}t_{0},$ which is hardly competitive with $2t_{0}$, and $%
N_{w}=30$.

In Fig.\ \ref{fig2} we report the local spin densities calculated by the LKF
under a high bias of $eV_{0}=0.4t_{0}$, spin components by the SVF,
translation method, and spin-orbit gauge method, for the pure Rashba case
with $t_{R}=0.1t_{0}$ in column (a), pure Dresselhaus case with $%
t_{D}=0.1t_{0}$ in column (c), and a mixed case with $%
(t_{R},t_{D})=(0.1,0.03)t_{0}$ in column (b). At the bottom of each column,
the local charge density obtained by the LKF with both a high bias of $%
eV_{0}=0.4t_{0}$ and a low bias of $eV_{0}=10^{-3}t_{0}$ is also reported.
At the top of each column, the spatially imaged spin vectors are from the
LKF results. Note that the LKF-based spin densities $\langle
S_{n}^{i}\rangle $ given by Eq.\ \eqref{spin density} have been normalized
by requiring $\left\vert \langle \mathbf{S}_{n=1}\rangle \right\vert =\hbar
/2$, while the spin components obtained by the quantum mechanical methods
are inherently of unit norm due to the normalized state kets.

Clearly in Fig.\ \ref{fig2} all the spin curves obtained by translation and
by spin-orbit gauge methods are identical to each other. These curves
further fit with those obtained by the LKF all quite well, except the
oscillating tails that appear in the LKF results. These oscillations result
from the nonequilibrium accumulation of the electron number that cannot be
taken into account in the quantum mechanical approaches. With low bias the
electrons behave like waves (as shown in the charge density curves in the
bottom panels of Fig.\ \ref{fig2}), which is the assumption in the quantum
mechanical approaches. The spin density curves obtained by translation or
spin-orbit gauge method match those obtained by the LKF with low bias
perfectly (not shown).

For the curves from SVFs, we use Eq.\ \eqref{SVF R} for the Rashba case of
Fig.\ \ref{fig2}(a), Eq.\ \eqref{SVF D} for the Dresselhaus case of Fig.\ %
\ref{fig2}(c), and Eq.\ \eqref{SVF} for the mixed case of Fig.\ \ref{fig2}%
(b). In region I of all three cases, the SVF curves match all the others
perfectly since in that region the orthogonality approximation Eq.\ %
\eqref{overlap} is in fact exact. Once the electron enters region II, the
error contained in Eq.\ \eqref{overlap} for the SVF curves starts to
accumulate. The error is still moderate in Figs.\ \ref{fig2}(a) and \ref%
{fig2}(c), but becomes drastic in the mixed case of Fig.\ \ref{fig2}(b), as
we have remarked previously in Sec.\ \ref{sec svf}.

Comparing the low-bias charge densities in all three cases, one can see that
an additional modulation appears in region II when both SOCs are present
[bottom panel of Fig.\  \ref{fig2}(b)]. This charge density modulation stems
from the angle-dependent spin-orbit potential, and will be explained in
detail later in Sec.\ \ref{sec charge density modulation}. 
\begin{figure*}[t]
\centering\includegraphics[width=\textwidth]{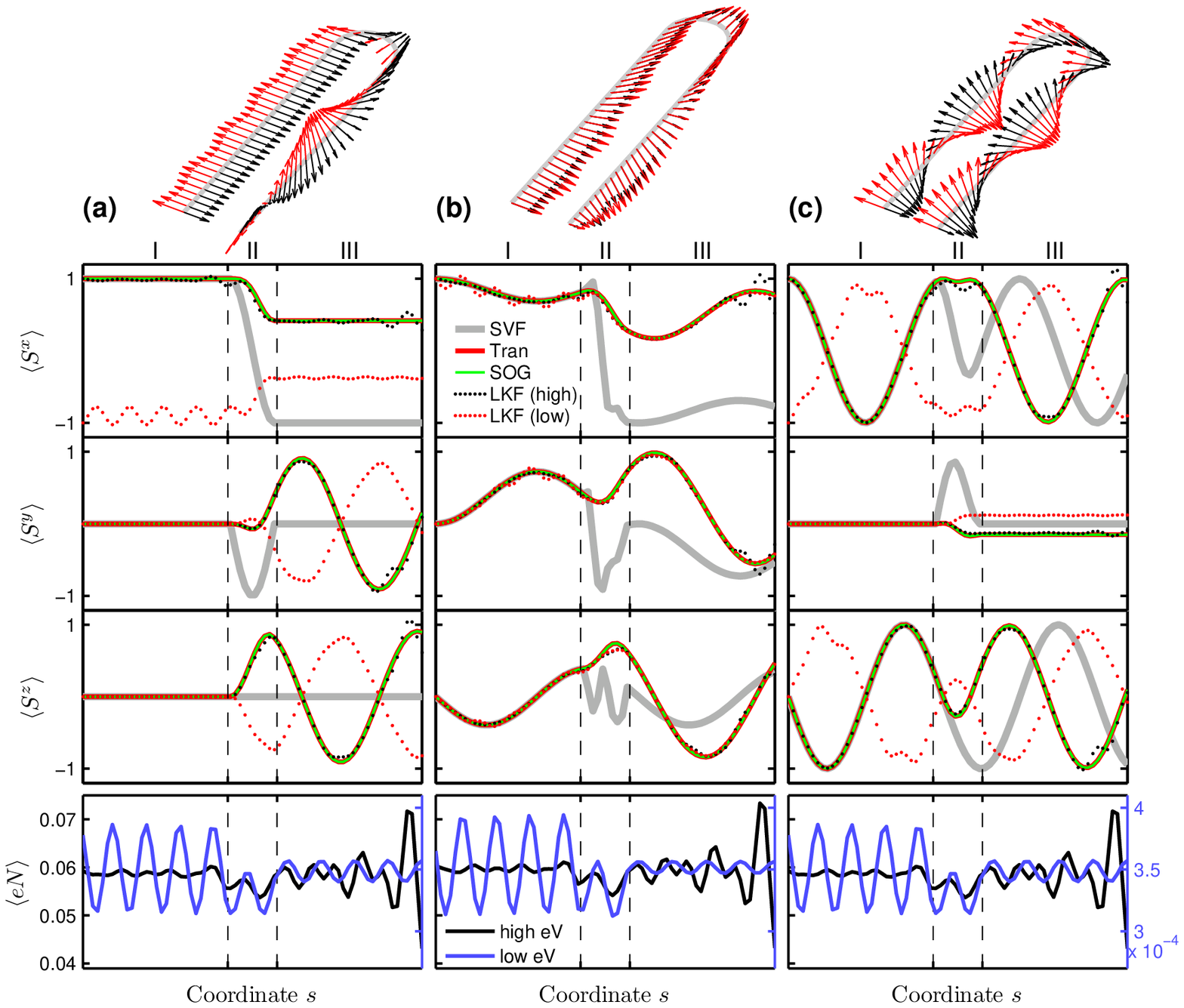}
\caption{(Color online) Similar to Fig.\ \protect\ref{fig2} but $N_{r}=10$.
Spin densities obtained by the LKF are shown for both low and high bias. In
the top subplots, black or red (gray) vectors correspond to high- or
low-bias LKF calculations.}
\label{fig3}
\end{figure*}

Before leaving for the stronger geometric potential case, we give a further
discussion over the Rashba channel: Fig.\ \ref{fig2}(a). Since $S^{x}$ is
one of the Rashba eigenstates in region I, the injected spin perfectly
retains its spin direction until the half-ring section is reached. Then
nonvanishing $\langle S^{z}\rangle $ component is induced in the curves from
the LKF, and translation or spin-orbit gauge method since the eigenstates of
the Rashba ring are no longer in-plane.\cite%
{Splettstoesser2003,Molnar2004,Frustaglia2004} The spin direction described
by the SVF, on the other hand, remains in-plane and perpendicular to the
transport, since after the orthogonal approximation [Eq.\ \eqref{overlap}]
the coplanar normal of a continuous 1D channel is still concluded as the
eigenstate. Hence the \textquotedblleft generalized
precessionless\textquotedblright\ transport in the curved 1D Rashba channel
predicted in Ref.\ \onlinecite{Liu2006b} may not work well. The $S^{x}$ spin
entering the half-ring region, in fact, starts to precess about the tilted
eigenstate of the Rashba ring with spin precession length $L_{so}$ [given
later in Eq.\ \eqref{Lso}], which matches exactly the period shown in $%
\langle S^{z}\rangle $ of Fig.\ \ref{fig2}(a). We will come back to this
tilted eigenstate later in Sec.\ \ref{sec spin precession special cases}.

\subsubsection{Strong geometric potential\label{sec strong Ug}}

Next we consider a U-channel with $N_{r}=10$ and the same $N_{w}$. The
geometric potential for such a $N_{r}$ is $\left\vert U_{g}\right\vert
=0.025t_{0},$ which is no longer negligible for the electrons. We keep the
same figure orientation as Fig.\ \ref{fig2}. The only information added in
Fig.\ \ref{fig3} is the spin components computed by the LKF with low bias.

The spin curves by translation and spin-orbit gauge methods are again
identical to each other and match the LKF curves with high bias well. SVF
curves this time become rather poor after entering the half-ring region
since there the change in the direction $\Delta \phi _{k}$ upon every
hopping is no longer small and the spinor overlap approximation %
\eqref{overlap} hardly applies.

For the spin curves obtained by the LKF with low bias, the effect of the
geometric potential $U_{g}$ can now be seen. The injected spin previously
parallel to the internal magnetic field direction $\mathbf{e}_{M}$ is here
reversed due to the reflection off the $U_{g}$ potential well and a certain
matching condition between the Fermi wavelength and the arm length $L$
within region I. Shifting the Fermi energy $E_{F}$, changing the length $L$,
or putting different SOC energies [such as Fig.\ \ref{fig3}(b)] will make
the reversal of the spin direction disappear. This is also why, in the
strong-bias regime, where a larger range of contributing states are
integrated, the reversal does not show.

The huge difference due to the stronger $U_{g}$ shown in Fig.\ \ref{fig3} is
hence only a special case: The reflection off the $U_{g}$ well and the
length matching happen to make the opposite spin state favored upon
injection. Here we conclude that the role played by the geometric potential
is merely a rather weak potential well that can be possibly sensed by the
electrons when the bending of the 1D structure is severe, and that even if $%
U_{g}$ is sensed, it serves simply as a potential well, which becomes
crucial only in the linear transport regime with a certain particular
matching condition between the Fermi wavelength and the channel size.

Apart from the spin behavior, the potential well nature of the $U_{g}$ can
be clearly identified by comparing the low-bias charge density curves shown
in the bottom panels of Figs.\ \ref{fig2} and \ref{fig3}. In the previous
weak $U_{g}$ case, the electron density distributes like a standing wave
from the source to the drain ends, while in the present strong $U_{g}$ case
the electron wave is disturbed by the central geometric potential well in
the half-ring part.

\subsection{Adiabatic and nonadiabatic transport regimes \label{sec spin
flip}}

\begin{figure}[t]
\centering\includegraphics[width=0.95\columnwidth]{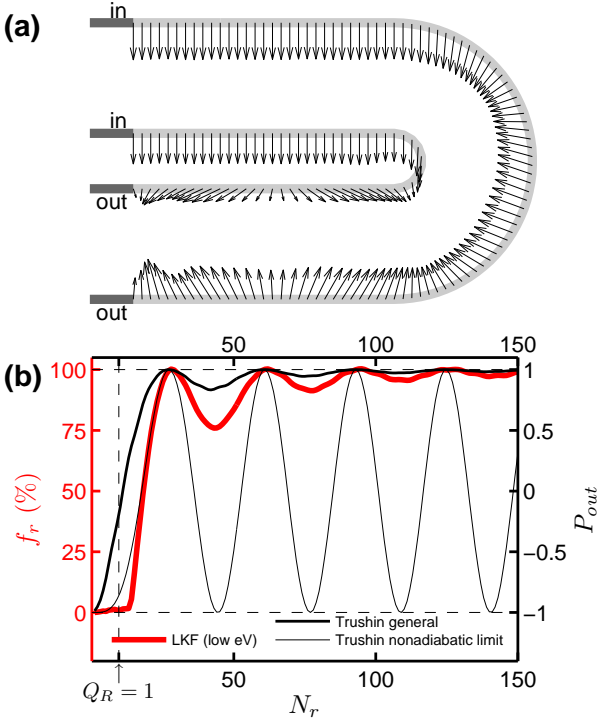}
\caption{(Color online) (a) Top view of the spin density vectors shown in
Figs.\ \protect\ref{fig2}(a) and \protect\ref{fig3}(a) with high bias. (b)
Spin flip ratio $f_{r}$ as a function of $N_{r}$ with $%
(t_{R},t_{D})=(0.1,0)t_{0}$ and the output spin polarization based on Ref.\ 
\onlinecite{Trushin2006}; the nonadiabatic curve is given by Eq.\ 
\eqref{Pout}.}
\label{fig4}
\end{figure}

Comparing further the previous two U-channels, one can see from the Rashba
cases that the spin initially injected at one of the Rashba wire eigenstates
may or may not follow the local eigenstate throughout the U-channel; see
Fig.\ \ref{fig4}(a). Certainly the key lies on the half-ring part, where the
larger the number $N_{r}$ is, the easier the spin can follow the eigenstate,
but the strength of the SOC is also an important factor.\cite{Frustaglia2004}

To make a quantitative investigation, we first define the following
spin-flip ratio,%
\begin{equation}
f_{r}\equiv \frac{1}{\pi }\cos ^{-1}\left( \frac{\mathbf{S}_{L}\cdot \mathbf{%
S}_{R}}{\left\vert \mathbf{S}_{L}\right\vert \left\vert \mathbf{S}%
_{R}\right\vert }\right) \frac{\left\vert \mathbf{S}_{R}\right\vert }{%
\left\vert \mathbf{S}_{L}\right\vert }\times 100\%,  \label{fr}
\end{equation}%
where $\mathbf{S}_{L}$ and $\mathbf{S}_{R}$ are the average spin direction
of the left and right arms, respectively, computed by the LKF. In the case
of the Rashba U-channel with $S^{x}$ injection, the spin is flipped from $%
+S^{x}$ to $-S^{x}$ when reaching the right arm, if the local eigenstate is
strictly followed. Thus the definition of Eq.\ \eqref{fr} helps us quantify
how well the local eigenstate is followed: whether a change in direction or
a shrink in the magnitude reduces the spin-flip ratio. Note that Eq.\ %
\eqref{fr} is also valid for pure Dresselhaus U-channels, provided that the
injected spin has to be oriented along $\pm y$, due to the $180%
{{}^\circ}%
$ turn of the U-channel. That is, as long as the spin transport is
sufficiently adiabatic, the injected spin is able to follow the local
eigenstate so that the spin is flipped after passing through the half-ring
region.

Let us first fix the Rashba SOC strength as $t_{R}=0.1t_{0}$ but change the
half-ring from small radius to larger ones, as shown in Fig.\ \ref{fig4}(b),
where a clear jump at about $N_{r}=16$ is observed. Within $N_{r}\lesssim 16$
the spin-flip ratio is nearly zero, showing that the spin can hardly follow
the local spin eigenstate when entering the half-ring region. At right side
of the jump, $f_{r}$ increases to 100\% and then exhibits a resonance-like
oscillation below the maximum value, in close analogy to Ref.\ %
\onlinecite{Trushin2006} and similar to some of the results reported in
Ref.\ \onlinecite{Zhang2007}. The oscillation period of about $32$
corresponds to a distance for the spin to complete a $2\pi $ of precession
angle under the Rashba SOC, i.e., two times the spin precession length,\cite%
{Nikolic2005a}%
\begin{equation}
L_{so}/a=\frac{\pi }{2}\frac{t_{0}}{t_{R}}.  \label{Lso}
\end{equation}%
Here we have $2L_{so}/a=10\pi $. This oscillation period can be well
described by Ref.\ \onlinecite{Trushin2006}, which we will discuss later. 
\begin{figure}[t]
\centering\includegraphics[width=0.95\columnwidth]{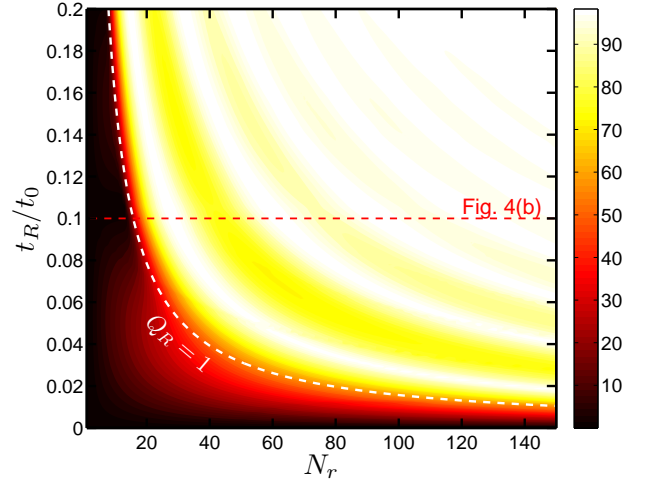}
\caption{(Color online) Spin-flip ratio $f_{r}$ defined in Eq.\ \eqref{fr}
as a function of $t_{R}$ and $N_{r}$, free of $t_{D}$. The white dashed line
given by $Q_{R}=1$ divides the transport into nonadiabatic (left-bottom) and
adiabatic (right-top) regimes. Horizontal dashed line corresponds to Fig.\ 
\protect\ref{fig4}(b).}
\label{fig5}
\end{figure}

The jump of $f_{r}$ and the wavelength of the resonance-like oscillation
depends on the SOC strength. Hence we next vary both Rashba strength and the
site number of the half-ring, and make the plot for $f_{r}$ as a function of 
$t_{R}$ and $N_{r}$ in Fig.\ \ref{fig5}. The $f_{r}$ pattern is clearly
divided into two regimes that can be perfectly described by the $Q_{R}=1$
curve, which was inspired by Ref.\ \onlinecite{Frustaglia2004}. The
adiabatic condition was previously argued as\cite{Stern1992,Aronov1993} $%
Q\gg 1$, where $Q=Q_{B}+Q_{R}$ includes the contribution from external
magnetic field and the Rashba field. In our analysis no external magnetic
field is applied, and the adiabatic condition reads $Q_{R}\gg 1$. The
definition of $Q_{R}$ from Ref.\ \onlinecite{Frustaglia2004} is here
reexpressed in terms of our tight-binding parameters, 
\begin{equation}
Q_{R}=\frac{2}{\pi }\frac{t_{R}}{t_{0}}N_{r},  \label{QR}
\end{equation}%
which implies that the increase of either $N_{r}$ or $t_{R}$ brings the
transport regime to adiabatic. Therefore the criterion that $Q_{R}\gg 1$
preserves the transport in the adiabatic regime is well agreed. Furthermore,
by a comparison with Eq.\ \eqref{Lso}, the meaning of $Q_{R}$ given by Eq.\ %
\eqref{QR} is transparent: $Q_{R}=N_{r}/\left( L_{so}/a\right) ,$ i.e., the
number of precession half-periods that the spin can complete within the
half-ring. Hence the condition $Q_{R}=1$ from our notation corresponds
exactly to the arc length of the half-ring that matches one spin precession
length $L_{so}$. The mathematical criterion for adiabatic transport, within
which the electron spins are able to follow the local field, means that the
electron spins have to be able to complete at least an angle of $\pi $ of
precession within the half-ring. One might attempt to extend this
interpretation to an arbitrary expanding angle of an arc, such as the
curvilinear QW considered in Ref.\ \onlinecite{Zhang2007}, but further
examination is left here as a possible extending work.

Finally, we compare our result with that of Ref.\ \onlinecite{Trushin2006},
which is shortly reviewed in the following. Trushin and Chudnovskiy
previously considered also a U-shaped 1D channel with Rashba SOC and solved
the transmission problem by matching the boundary conditions.\cite%
{Trushin2006} A spin polarization was defined as $%
P=(j^{+}-j^{-})/(j^{+}+j^{-})$, where $j^{\pm }$ is the probability current
of the $\pm $ eigenspin components. The polarized wave occupying the $+$
Rashba eigenstate was assumed as the incoming state so that $P_{in}=1$ and
the spin polarization for the outgoing wave $P_{out}$ is the main quantity
of interest. If the injected spin remains at its local eigenstate, $P_{out}=1
$ is expected, which is the adiabatic limit. Oppositely, a strong
nonadiabatic limit leads to a simplified expression,\cite{Trushin2006}%
\begin{equation}
P_{out}=\cos \left( \pi Q_{R}\sqrt{1+Q_{R}^{-2}}\right) ,  \label{Pout}
\end{equation}%
which has been translated to our tight-binding language. As shown in Fig.\ %
\ref{fig4}(b), the oscillation matches our result; the curves in the low $%
N_{r}$ region ($Q_{R}$ small) match particularly well. For larger $Q_{R}$,
Eq.\ \eqref{Pout} approaches $\cos (\pi Q_{R})$ with the oscillation period $%
\Delta Q_{R}=2=\Delta N_{r}/(L_{so}/a)$, which well describes the period of $%
\Delta N_{r}=2L_{so}/a$ in Fig.\ \ref{fig4}(b), in agreement with our
previous discussion. For general $Q_{R}$, $P_{out}$ can be computed
following their results and is plotted also in Fig.\ \ref{fig4}(b) (with $%
t_{R}=0.1t_{0}$ and $E_{F}=0.2t_{0}$). Overall, the oscillation behavior in
both the nonadiabatic and general cases from Ref.\ \onlinecite{Trushin2006}
agrees with our result.

In the above discussion, we have injected an $S^{x}$ spin, which is the
eigenstate of the Rashba wire. The fact that the eigenstates in the ring
differ from those in the wire by a tilt angle is the origin that a spin
starting with $S^{x}$ in the U-channel can never perfectly follow the local
eigenstate in the half-ring part. In principle, when the spin direction
happens to match the ring eigenstate when the electron is just about to
enter the ring, the local ring eigenstate can then be well followed. In the
following section, we will show that these special cases do exist, provided
that the length $L$ and the orientation angle of the injected spin are
precisely designed.

\subsection{More on charge and spin transport\label{sec charge and spin
transport}}

The last goal to be carried out here is to reveal some of the interesting
transport properties, regarding both charge and spin, based on quantum
transport calculations.

\subsubsection{Charge density modulation: Emergence of spin-orbit potential 
\label{sec charge density modulation}}

\begin{figure}[t]
\centering\includegraphics[width=\columnwidth]{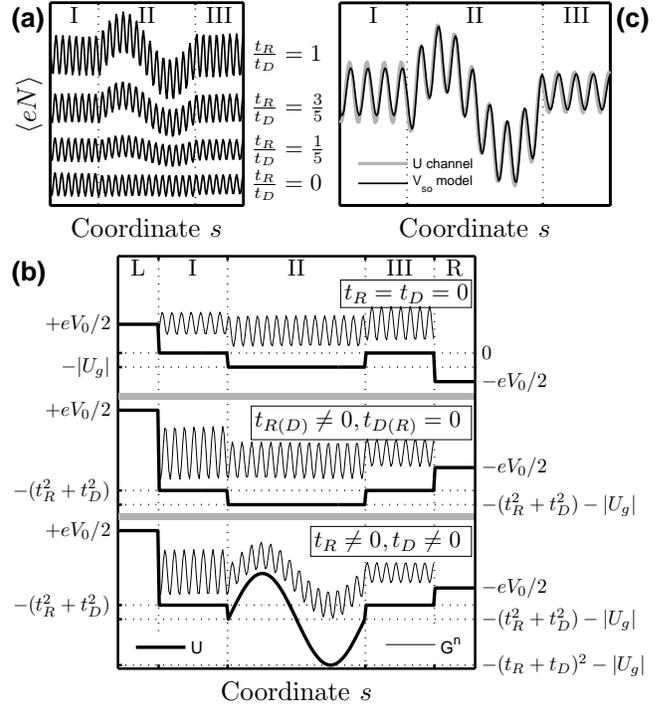}
\caption{(a) Formation of the charge density modulation in a $\left(
N_{w},N_{r}\right) =\left( 50,100\right) $ U-channel with $%
eV_{0}=10^{-3}t_{0}$, $t_{D}$ fixed at $0.02t_{0}$, and $t_{R}$ varied from $%
0$ to $t_{D}$. (b) 1D Schr\"{o}dinger problem for a 1D linear chain subject
to the spin-orbit potential model. The top, middle, and bottom panels
account for the zero SOC, single type of SOC, and mixed type of SOCs,
respectively. Parameters used here are identical to the U-channel shown in
(a). In addition to regions I, II, and III that correspond to those in the
U-channel, the left (L) and right (R) leads are also labeled. (c) Local
charge density $\langle eN\rangle $ for the U-channel and the electron
correlation function $G^{n}\propto \left\vert \protect\psi \left( s\right)
\right\vert ^{2}$ for the linear chain, both considering $%
t_{R}=t_{D}=0.02t_{0}$. Calculations here are all with $E_{F}=0.2t_{0}$.}
\label{fig6}
\end{figure}

As previously remarked in Sec.\ \ref{sec weak Ug} [or specifically the
bottom panel of Fig.\ \ref{fig2}(b)], an additional modulation to the
low-bias charge density in the half-ring region appears when both terms of
SOCs are present. In Fig.\ \ref{fig6}(a), we show the formation of this
charge density modulation in a $\left( N_{w},N_{r}\right) =\left(
50,100\right) $ U-channel by fixing $t_{D}=0.02t_{0}$ and varying from $%
t_{R}=0$ to $t_{R}=t_{D}$, with $eV_{0}=10^{-3}t_{0}$ and $E_{F}=0.2t_{0}$.
Clearly, the modulation appears only when $t_{R}t_{D}\neq 0$ and reaches its
maximum when both SOCs are of the same strength. This modulation was
similarly obtained in a recent study of the anisotropic spin transport in
mesoscopic rings,\cite{Wang2008} but the origin there was not clear. In the
following we provide a simple quantum mechanical picture to account for this
modulation.

Recall the anisotropic spin splitting, Eq.\ \eqref{zeta}. By solving for $%
k_{\pm }$ from $E_{F}=\hbar ^{2}k_{\pm }^{2}/2m\pm \zeta k_{\pm }$ and
defining the Fermi wave vector as $k_{F}=\left( k_{+}+k_{-}\right) /2$, we
have%
\begin{equation}
k_{F}a=\sqrt{\frac{E_{F}}{t_{0}}+\left( \frac{t_{so}}{t_{0}}\right) ^{2}},
\label{kFa}
\end{equation}%
where%
\begin{equation}
t_{so}\left( \phi _{k}\right) =\frac{\zeta }{2a}=\sqrt{%
t_{R}^{2}+t_{D}^{2}-2t_{R}t_{D}\sin 2\phi _{k}}.  \label{tso}
\end{equation}%
The fact that the Fermi wavelength $\lambda _{F}=2\pi /k_{F}$ is forced
(when $t_{R}t_{D}\neq 0$) to be modulated upon changing propagation angle $%
\phi _{k}$ can be mapped to a \emph{linear 1D chain} with a
position-dependent local potential. Defining%
\begin{equation}
\frac{V_{so}}{t_{0}}=-\left( \frac{t_{so}}{t_{0}}\right) ^{2},  \label{Vso}
\end{equation}%
Eq.\ \eqref{kFa} then reads $k_{F}a=\sqrt{\left( E_{F}-V_{so}\right) /t_{0}}$%
, as if the electron were propagating in a 1D linear system subject to a
potential $V_{so}$, i.e., the electron is governed by $\left(
p^{2}/2m+V_{so}\right) \psi =E\psi $. This interpretation is exact when $%
V_{so}$ is a constant potential; when $V_{so}$ is position dependent but
weak compared with $E_{F}$, the argument is still a good approximation.

We therefore consider a 1D linear chain subject to, together with the
geometric potential, the full local potential,%
\begin{equation}
U\left( s\right) =%
\begin{cases}
V_{so}\left( +\pi /2\right) , & s\in \left[ 0,L\right]  \\ 
V_{so}\left( \phi _{k}\right) +U_{g}, & s\in \left[ L,L+\pi r\right]  \\ 
V_{so}\left( -\pi /2\right) , & s\in \lbrack L+\pi r,2L+\pi r]%
\end{cases}%
,  \label{U}
\end{equation}%
where $\phi _{k}$ as a function of $s$ is taken identical to Eq.\ \eqref{phi}%
. The 1D Schr\"{o}dinger problem subject to the potential $U(s)$ given by
Eq.\ \eqref{U}, $\left[ -\hbar ^{2}\partial ^{2}/2m\partial s^{2}+U(s)\right]
\psi (s)=E\psi (s)$, can be analytically cumbersome due to the irregular
shape of $U(s)$, but can be easily solved by the quantum transport formalism
introduced in Sec.\ \ref{sec lkf} at an even lower cost. The lead
self-energy, Eq.\ \eqref{self-energy}, with $t_{M}=0$ is taken for both left
and right leads (with potentials $+eV_{0}/2$ and $-eV_{0}/2$, respectively)
to simulate the incoming and outgoing waves. The squared norm of the wave
function in such an equilibrium problem corresponds to the electron
correlation function $G^{n}\left( E\right) $ (equivalent to $-iG^{<}$) that
can be obtained from $f_{0}\left( E\right) A\left( E\right) $.\cite%
{Datta1995} The Fermi function $f_{0}$ will be taken as unity, concerning
the currently assumed zero temperature, and the spectral function $A$ can be
obtained from $A=G^{R}\Gamma G^{A}$, where $G^{R}$ and $G^{A}$ are the
retarded and advanced Green's function of the linear chain, respectively;
the broadening function is given by $\Gamma =i(\Sigma _{L}-\Sigma _{L}^{\dag
})$. Note that we have turned off the contribution of the right lead to the
broadening function to suppress the inflow of the particles from the right
leads. This is equivalent to setting the wave function in the outgoing
region as $\left. \psi \left( s\right) \right\vert _{s\geq L+\pi r}\propto
e^{+iks}$, which is a usual assumption taken in most quantum physics
textbooks.

The potential profile $U\left( s\right) $ together with $G^{n}\propto
\left\vert \psi \left( s\right) \right\vert ^{2}$ is plotted in Fig.\ \ref%
{fig6}(b) for three situations: zero SOC, single type of SOC, and mixed type
of two SOCs. The last case clearly resembles the charge density modulation
in the U-channel with $t_{R}t_{D}\neq 0$, and the present spin-orbit
potential model seems to work well. Thus the modulation of the electron
density profile simply reflects the position-dependent spin-orbit potential,
Eq.\ \eqref{Vso}, that is usually small compared with $E_{F}$. Indeed, in
Fig.\ \ref{fig6}(c) we further compare the local charge density $\langle
eN\rangle $ in the U-channel with the electron correlation function $G^{n}$
in the linear chain. The difference between them is only up to a tiny phase
shift. Detailed parameters used here in Fig.\ \ref{fig6} are given in the
caption thereof.

When $\left\vert V_{so}\right\vert $ is close to $E_{F}$, either by
strengthening the SOC parameters or lowering the Fermi energy, the phase
shift grows, but the $G^{n}$ calculated for the linear chain and the $%
\langle eN\rangle $ calculated for the U-channel still are similar in their
shapes (not shown). The present model hence works equally well to explain
the charge density modulation, which we conclude as originating from the
emergence of the angle-dependent spin-orbit potential when $t_{R}t_{D}\neq 0$%
.

\subsubsection{Spin precession in special cases\label{sec spin precession
special cases}}

In this subsection we discuss spin precessions in three special cases. In
the first two cases, only one type of SOC is considered, and we inject the
spin oriented as the theoretically predicted eigenstate for the \emph{ring }%
(which is different from that for the wire), and adjust $t_{R}$ or $t_{D}$
such that the length $L$ equals exactly two times the spin precession length 
$L_{so}$. The injected spin arriving at the half-ring returns exactly to the
eigenspin direction, such that the previously derived tilted eigenstate,
e.g., the eigenspinor for clockwise-propagating $\downarrow $ eigenspinor in
the Rashba ring from Ref.\ \onlinecite{Frustaglia2004},%
\begin{equation}
\chi _{R-}^{\downarrow }\left( \phi \right) =\left( 
\begin{array}{c}
\sin \left( \gamma _{R}/2\right)  \\ 
e^{i\phi }\cos \left( \gamma _{R}/2\right) 
\end{array}%
\right) =\left( 
\begin{array}{c}
\cos [(\pi -\gamma _{R})/2] \\ 
e^{i\phi }\sin [(\pi -\gamma _{R})/2]%
\end{array}%
\right) ,  \label{chi down R}
\end{equation}%
with the tilt angle $\gamma _{R}=\tan ^{-1}Q_{R}$, can be numerically
examined. Note that here $\phi $ is the azimuthal angle in the polar
coordinate ($\phi =0$ defined along $+x$ axis), rather than the propagation
angle $\phi _{k}$.

\begin{figure}[t]
\centering\includegraphics[width=\columnwidth]{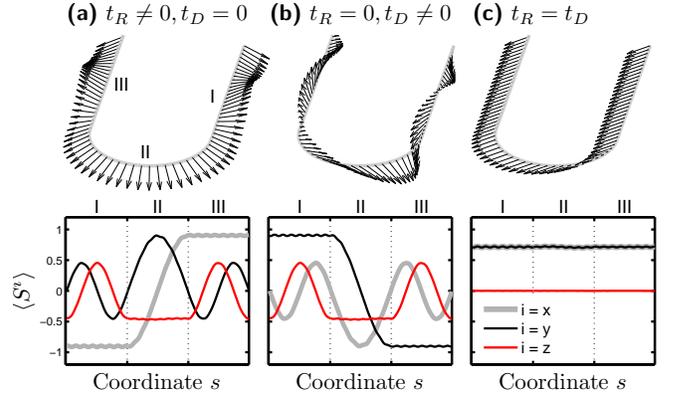}
\caption{(Color online) Spin precession in a $\left( N_{w},N_{r}\right)
=\left( 50,50\right) $ U-channel with (a) $\left( t_{R},t_{D}\right) =\left( 
\protect\pi /N_{w},0\right) t_{0}$, (b) $\left( t_{R},t_{D}\right) =\left( 0,%
\protect\pi /N_{w}\right) t_{0}$, and (c) $t_{R}=t_{D}=\left( \protect\pi /%
\protect\sqrt{2}N_{w}\right) t_{0}$. The injected spin is oriented in (a) as
the Rashba ring eigenstate Eq.\ \eqref{chi down R}, in (b) as the
Dresselhaus ring eigenstate Eq.\ \eqref{chi down D}, and in (c) as the
persistent spin helix eigenstate Eq.\ \eqref{chi down ReD}. Note that in the
imaging of the space-resolved spin vectors, only half of the vectors are
drawn in order for clarity.}
\label{fig7}
\end{figure}

To account for the two extreme cases of pure Rashba and pure Dresselhaus
using one single formula, we write the clockwise-propagating $\downarrow $
eigenspinor as%
\begin{equation}
\chi _{-}^{\downarrow }\left( \phi \right) =\left( 
\begin{array}{c}
\cos [(\pi -\gamma )/2] \\ 
e^{i\varphi }\sin [(\pi -\gamma )/2]%
\end{array}%
\right) ,  \label{chi down}
\end{equation}%
where $\varphi $ is given, similar to Eq.\ \eqref{vphi}, as%
\begin{equation}
\varphi =\arg \left[ t_{R}\cos \phi -t_{D}\sin \phi +i\left( t_{R}\sin \phi
-t_{D}\cos \phi \right) \right] .  \label{vphi tR tD}
\end{equation}%
The tilt angle in Eq.\ \eqref{chi down} is given by%
\begin{equation}
\gamma =\tan ^{-1}\left( \sqrt{Q_{R}^{2}+Q_{D}^{2}}\right) ,
\label{gamma general}
\end{equation}%
where $Q_{D}$ is defined similar to Eq.\ \eqref{QR} as $Q_{D}=(2/\pi
)(t_{D}/t_{0})N_{r}.$ Equation \eqref{chi down} clearly recovers the Rashba
ring case of Eq.\ \eqref{chi down R} since $\left. \varphi \right\vert
_{t_{D}=0}=\phi $ from Eq.\ \eqref{vphi tR tD} and $\left. \gamma
\right\vert _{t_{D}=0}=\gamma _{R}$ from Eq.\ \eqref{gamma general}, can
cover the $t_{R}=0$ Dresselhaus ring case,%
\begin{equation}
\chi _{D-}^{\downarrow }=\left( 
\begin{array}{c}
\cos [(\pi -\gamma _{D})/2] \\ 
e^{-i(\pi /2+\phi )}\sin [(\pi -\gamma _{D})/2]%
\end{array}%
\right) ,  \label{chi down D}
\end{equation}%
where $\gamma _{D}=\tan ^{-1}Q_{D}$, but does not apply for general $%
t_{R}t_{D}\neq 0$ cases.

The last case is $t_{R}=t_{D}$, corresponding to the persistent spin-helix
in 2DEG,\cite{Bernevig2006,Liu2006c,Koralek2009} the main feature of which
is the fixed eigenspin directions. In the present case, the $\downarrow $
eigenspinor is%
\begin{equation}
\chi _{R=D}^{\downarrow }=\frac{1}{\sqrt{2}}\left( 
\begin{array}{c}
e^{-i\pi /4} \\ 
1%
\end{array}%
\right) ,  \label{chi down ReD}
\end{equation}%
corresponding to $E_{k}=\hbar ^{2}k^{2}/2m-\zeta k$. Whether this
eigenstate, valid for 2DEG, still works in the curved 1D system, is what we
are about to answer.

(a) \textit{Rashba ring eigenstate.} We begin with $t_{D}=0$ and inject $%
\chi _{R-}^{\downarrow }\left( \phi =\pi \right) $ as given by Eq.\ %
\eqref{chi down R} in a U-channel with $\left( N_{w},N_{r}\right) =\left(
50,50\right) $. We tune $t_{R}/t_{0}=\pi /N_{w}\approx 0.063$ such that $%
N_{w}=\pi t_{0}/t_{R}=2L_{so}/a$ [see Eq.\ \eqref{Lso}] ensures the return
of the injected spin to its initial spin direction, which is the eigenstate
of the ring at $\phi =\pi $, after going through region I. As shown in Fig.\ %
\ref{fig7}(a), the spin entering the half-ring region remains perfectly in
the eigenstate. (Note that we have chosen another view angle to focus on the
half-ring part; injection conditions remain the same as in previous
discussions.) The curves of $\langle S^{i}\rangle $ within region II can be
well described by $\langle \chi _{R-}^{\downarrow }\left( \phi \right) |\vec{%
\sigma}|\chi _{R-}^{\downarrow }\left( \phi \right) \rangle $ [given below
in Eq.\ \eqref{<s> tilted}] with $\phi $ running from $\pi $ to $0$. The
validity of the previously derived tilted eigenstate in Rashba rings is
hence numerically verified.

Note the subtle difference between the special spin injection here and in
Sec.\ \ref{sec spin flip}, where we injected an inplane $S^{x}$: an
eigenstate of the wire. The precise design of the length $L$ and the
orientation of the injected spin allows the spin to stay perfectly in the
tilted ring-eigenstate. The $x$-component of spin with such a precise design
can always be flipped, but should be regarded as a special situation.

(b) \textit{Dresselhaus ring eigenstate.} We continue with $t_{R}=0$ and
inject $\chi _{D-}^{\downarrow }\left( \phi =\pi \right) $ given by Eq.\ %
\eqref{chi down D}. In this case we similarly have $t_{D}/t_{0}=\pi /N_{w}$.
Again the spin arriving at the half-ring enters its eigenstate and remains
so until leaving the ring, as shown in Fig.\ \ref{fig7}(b). The spin
components $\langle S^{i}\rangle $ within region II can be well described by 
$\langle \chi _{D-}^{\downarrow }\left( \phi \right) |\vec{\sigma}|\chi
_{D-}^{\downarrow }\left( \phi \right) \rangle $, and the validity of the
Dresselhaus ring eigenstate is also numerically verified. For both Figs.\ %
\ref{fig7}(a) and \ref{fig7}(b), spin components from Eq.\ \eqref{chi down},%
\begin{equation}
\langle \chi _{-}^{\downarrow }\left( \phi \right) |\vec{\sigma}|\chi
_{-}^{\downarrow }\left( \phi \right) \rangle =\left( 
\begin{array}{c}
\sin \left( \pi -\gamma \right) \cos \varphi  \\ 
\sin \left( \pi -\gamma \right) \sin \varphi  \\ 
\cos \left( \pi -\gamma \right) 
\end{array}%
\right) ,  \label{<s> tilted}
\end{equation}%
describes the $\langle S^{i}\rangle $ curves in region II well.

We remind the reader here that the eigenstate given by Eq.\ \eqref{chi down}
is intended only for the two extreme cases of $t_{R(D)}\neq 0,t_{D(R)}=0$
discussed above, although a solution given in Ref.\ \onlinecite{Wang2005},
similar to our Eq.\ \eqref{chi down}, was claimed to be valid for rings in
the presence of both Rashba and Dresselhaus terms. The simple reason why the
form of Eq.\ \eqref{chi down} does not apply for general cases of $%
t_{R}t_{D}\neq 0$ is that the tilt angle $\gamma $ [Eq.\ 
\eqref{gamma
general}] does not recover $\pi /2$ when the $t_{R}=t_{D}\neq 0$ persistent
spin helix state is reached, which is true as we will next numerically show.

(c) \textit{Persistent spin-helix eigenstate.} We proceed by considering $%
t_{R}=t_{D}=(\pi /\sqrt{2}N_{w})t_{0}\approx 0.044t_{0}$, keeping the size
of the U-channel unchanged. The injected spin state is oriented as $\chi
_{R=D}^{\downarrow }$, given in Eq.\ \eqref{chi
down ReD}. As expected, the injected spin stays at this eigenstate in region
I, as shown in Fig.\ \ref{fig7}(c). Somewhat surprisingly, however, the
injected spin remains precessionless throughout the whole U-channel, even in
the half-ring region. Therefore, the persistent spin-helix eigenstate,
originally derived for a 2DEG, is equally valid in straight wires and curved
rings. This is in sharp contrast to the pure Rashba and pure Dresselhaus
cases, for which the eigenstates in wires and in rings are different.

\section{Experimental aspects\label{sec exp}}

The U-shaped 1D channel theoretically discussed in the present paper can be
experimentally prepared by using AFM lithography,\cite{Ihn2010} i.e., local
oxidation\cite{Fuhrer2002} written by an AFM tip on the sample. The oxide
lines turn out to completely deplete the 2DEG underneath, and can hence
confine the electron gas in the desired nanostructure. A schematic sketch of
the U-channel based on this technique is suggested in Fig.\ \ref{fig1}(c).
According to the present fabrication ability (see, for example, Ref.\ %
\onlinecite{Fuhrer2001}), however, the ring radius is of the order of $100%
\unit{nm}$, and the induced geometric potential is rather weak: $10^{-2}$ $%
\func{meV}$ for GaAs-based quantum wells. To focus on the effect of the
geometric potential, a single turn as in our U-channel is not enough; a
series of geometric potential wells such as a sinelike waveguide similar to
the design reported in a recent experiment on photonic crystal\cite%
{Szameit2010} may give rise to a resonance that could potentially be
measured.

The spin injection assumed here may be realized either electrically or
optically. The former requires a ferromagnetic source contact and may
further complicate the sample fabrication and even the transport properties.
The latter, optical spin injection, has been mature in generating spin
packets that can be electrically manipulated.\cite{Flatt'e2007} Regarding
the U-channel sketched in Fig.\ \ref{fig1}(c), the adiabatic-nonadiabatic
spin transport discussed here may be experimentally tested by optically
pumping at the source a spin packet that can be electrically dragged to the
drain end by applying a bias voltage between the source and the drain
contacts. Optical spin detection of the spin packet at the drain end shows
whether the spin is reversed (adiabatic), is decayed (spin relaxed), or
remains (nonadiabatic), compared with the injected spin direction. The laser
spot size for the optical spin injection or detection, typically of a few
hundreds of microns, may impose a corresponding limit on the design but
could be possibly overcome by hard masks. A top gate covering the U-channel
may control the Rashba SOC strength\cite{Nitta1997,Grundler2000,Bergsten2006}
and switch the transport regimes between adiabatic and nonadiabatic,
provided that the effect of the Dresselhaus term is well treated.

Experimental proof of the interesting charge and spin transport properties
discussed in Sec.\ \ref{sec charge and spin transport} are also expected. In
particular, the charge density modulation in the presence of both Rashba and
linear Dresselhaus (001) SOCs discussed in Sec.\ \ref{sec charge density
modulation} requires measurement on the local charge densities only and
should be possible. The profile of the charge density modulation simply
reflects the angle-dependent spin-orbit potential and hence determines the
type of the SOCs: flat for the single type of SOC and sinelike for the mixed
type of SOC. Note that in our discussion of the U-channel, the nature of $%
\sin 2\phi $ dependence of the spin-orbit potential provides the modulation
in the half-ring with one period of the sine function. In the case of a full
ring, we expect two periods then.

An alternative to preparing a 1D channel is the V-groove QW based on
electron beam lithography,\cite{Ihn2010} but so far application of this
technique to curved 1D QWs has not been seen.

\section{Conclusion\label{sec conclusion}}

In conclusion, we have rederived the Hamiltonian for a curved 1D structure
in the presence of SOC. Applied to the 1D ring, the Hamiltonian is not only
consistent with the previously proposed proper Hamiltonian for a Rashba ring,%
\cite{Meijer2002} but also contains a curvature-induced geometric potential,
which was first derived in Ref.\ \onlinecite{Costa1981}, but less discussed
in the literature of ring issues. The U-shaped 1D channel is further taken
to be a specific example to investigate the role of this geometric
potential, as well as to compare the spin densities obtained by the LKF with
the previous quantum mechanical approaches. Both translation\cite{Liu2006b}
and spin-orbit gauge\cite{Chen2008a} methods mostly agree with the LKF
results, even though the underlying assumption is rather simplified: to
assume an ideally injected spin at the initial position, and the technique
is rather artificial: to drag the injected spin by operating quantum
operators. Whether the SVF,\cite{Liu2006b} a further approximating result
from the translation method, may work well or not depends on whether the
orthogonality approximation [Eq.\ \eqref{overlap}] is valid enough: exact
for a straight 1D structure, of moderate error for a ring with a single type
of SOC, and poor for a ring with a mixed type of SOCs.

The influence of the geometric potential taken into account in the LKF
calculation is shown to be sensible only when the turn of the channel is
sharp and the transport is under low bias. Overall the role played by the
geometric potential is moderate, just like a local potential well, and can
be drastic [such as the reversal of the injected spin state shown in Fig.\ %
\ref{fig3}(a) or \ref{fig3}(c)] only when a certain resonance condition is
reached.

We have also discussed the spin transport in adiabatic and nonadiabatic
regimes. In addition to the increase of the geometric potential when making
the turn sharper by reducing the number of site $N_{r}$ in the turning part,
the transport becomes nonadiabatic since the change of the local eigenstate
becomes rapid. The spin transport shows adiabatic behavior when the turn is
smooth and the SOC is strong enough, which agrees with the previously stated
adiabatic condition.\cite{Stern1992,Aronov1993,Frustaglia2004} We have also
compared our results with a recent similar work by Trushin and Chudnovskiy,%
\cite{Trushin2006} which showed good agreement.

The last part of the numerical results revealed interesting charge and spin
transport properties. For charge transport, the interplay between the Rashba
and linear Dresselhaus (001) SOCs leads to anisotropic spin splitting, and
hence an angle-dependent Fermi wavelength. Charge transport in a curved 1D
chain subject to a modulated Fermi wavelength is therefore mapped to
transport in a 1D linear chain subject to a position-dependent potential. We
have shown that the charge density modulation that appears only when $%
t_{R}t_{D}\neq 0$ in the half-ring part of the U-channel can be well
explained by the spin-orbit potential model. For spin transport, we have
shown spin precession patterns in three special cases, which are equivalent
to numerically testing the validity of the previously predicted tilted
eigenstates of the Rashba rings and Dresselhaus rings, as well as that of
the persistent spin-helix state.

\begin{acknowledgments}
We appreciate the National Science Council of Taiwan, Grant No.\ NSC
98-2112-M-002-012-MY3, for supporting the former part of this work. M.H.L.\
acknowledges the Alexander von Humboldt Foundation for supporting the later
part of this work, and is grateful to C.\ H.\ Back for sharing his
experimental aspects and K.\ Richter for valuable suggestions.
\end{acknowledgments}

\bibliographystyle{apsrev4-1}
\bibliography{mhl2}

\end{document}